\documentclass[12pt]{article}
\usepackage{amsmath,amssymb,bbm,stmaryrd,graphicx}

\def\slashchar#1{\setbox0=\hbox{$#1$}           % set a box for #1
   \dimen0=\wd0                                 % and get its size
   \setbox1=\hbox{/} \dimen1=\wd1               % get size of /
   \ifdim\dimen0>\dimen1                        % #1 is bigger
      \rlap{\hbox to \dimen0{\hfil/\hfil}}      % so center / in box
      #1                                        % and print #1
   \else                                        % / is bigger
      \rlap{\hbox to \dimen1{\hfil$#1$\hfil}}   % so center #1
      /                                         % and print /
   \fi}
   
\def\ad{\text{ad}}

\def\tr{\text{tr}}

\def\hV{\hat {\cal V}}

\def\writecenter#1{                     
   \rlap{\hbox to 50mm{\hfil#1\hfil}}   %center #1
   }

\def\be{\begin{equation}}
\def\ee{\end{equation}}
\def\ben{\begin{displaymath}}
\def\een{\end{displaymath}}
\def\bea{\begin{eqnarray}}
\def\eea{\end{eqnarray}}

\def\ft#1#2{{\textstyle {\frac{#1}{#2}} }}

\newcommand{\leftb}{|\![\,}
\newcommand{\rightb}{]\!|\,}

\def\mua{\mu}
\def\mub{\nu}

\def\cMa{{\cal M}}
\def\cMb{{\cal N}}
\def\cMc{{\cal P}}
\def\cMd{{\cal Q}}

\def\cAa{{\cal A}}
\def\cAb{{\cal B}}
\def\cAc{{\cal C}}

\makeatletter \@addtoreset{equation}{section} \makeatother

\begin{document}

\begin{titlepage}
\begin{center}

\hfill {\tt ENSL-00147449}\\

\vskip 1.5cm 
\begin{center}
{\large {\bf GAUGING HIDDEN SYMMETRIES}}\\[2ex]
{\large {\bf IN TWO DIMENSIONS}}
\end{center}

\vskip .7cm

{\bf Henning Samtleben$^{1}$ and Martin Weidner$^{2}$} \\

\vskip 22pt

$^{1}\;${\em Laboratoire de Physique\\[-.6ex]
Ecole Normale Sup{\'e}rieure Lyon\\[-.6ex]
46, all\'ee d'Italie\\[-.6ex]
F-69364 Lyon, CEDEX 07, France}

\vskip 8pt

$^{2}\;${\em Department of Economics\\[-.6ex]
University of Southern California\\[-.6ex]
3620 S. Vermont Ave. KAP 300\\[-.6ex]
Los Angeles, CA 90089, U.S.A.}\\

\vskip 15pt

{{\tt henning.samtleben@ens-lyon.fr, mweidner@usc.edu}} \\

\vskip 0.8cm

\end{center}

\vskip 8mm

\begin{center} {\bf ABSTRACT}\\[3ex]

\begin{minipage}{15cm}
\small
We initiate the systematic construction of 
gauged matter-coupled supergravity theories in two dimensions.
Subgroups of the affine global symmetry group of toroidally compactified supergravity
can be gauged by coupling vector fields with minimal couplings and a particular topological term.
The gauge groups typically include hidden symmetries that are not among the target-space isometries
of the ungauged theory.
The gaugings constructed in this paper are described group-theoretically in terms of a constant 
embedding tensor subject to a number of constraints which parametrizes the different theories
and entirely encodes the gauged Lagrangian.

The prime example is the bosonic sector of the maximally supersymmetric theory
whose ungauged version admits an affine $\mathfrak{e}_{9}$ global symmetry algebra.
The various parameters (related to higher-dimensional $p$-form fluxes, geometric and
non-geometric fluxes, etc.) 
which characterize the possible gaugings, combine into an embedding tensor transforming in
the basic representation of~$\mathfrak{e}_{9}$.
This yields an infinite-dimensional class of maximally supersymmetric theories in
two dimensions. We work out and discuss several examples of higher-dimensional origin
which can be systematically
analyzed using the different gradings of~$\mathfrak{e}_{9}$.

\end{minipage}
\end{center}
\noindent

\vfill

May 2007

\end{titlepage}

\pagenumbering{roman}

\tableofcontents

\newpage
\pagenumbering{arabic}
\setcounter{page}{1}

%%%%%%%%%%%%%%%%%%%%%%%%%%%%%%%%%%%%%%%%%%%%%%%%%%%%%%%%%%%%%%%
\section{Introduction}
%%%%%%%%%%%%%%%%%%%%%%%%%%%%%%%%%%%%%%%%%%%%%%%%%%%%%%%%%%%%%%%

One of the most intriguing features of extended supergravity theories is the exceptional global
symmetry structure they exhibit upon dimensional reduction~\cite{Cremmer:1979up}.
Eleven-dimensional supergravity when compactified on a $d$-torus $T^{d}$ gives rise to
an $(11\!-\!d)$-dimensional maximal supergravity with the exceptional global symmetry group ${\rm E}_{d(d)}$ 
and Abelian gauge group $U(1)^{q}$, where $q$ is the dimension of some (typically irreducible)
representation of ${\rm E}_{d(d)}$ in which the vector fields transform.
The only known supersymmetric deformations of these theories are the so-called gaugings in 
which a (typically non-Abelian) subgroup of ${\rm E}_{d(d)}$ is promoted to a local gauge group
by coupling its generators to a subset of the $q$ vector fields.
The resulting theories exhibit interesting properties such as mass-terms for the fermion fields 
and a scalar potential that provides masses for the scalar fields and may support de Sitter and
Anti-de Sitter ground states of the theory~\cite{deWit:1982ig}. 
Recently, gauged supergravities have attracted particular interest in the context of non-geometric and
flux compactifications~\cite{Grana:2005jc} where they describe the resulting low-energy effective theories
and in particular allow to compute the effective scalar potentials induced by particular flux configurations.

A systematic approach to the construction of gauged supergravity theories has been set up
with the group-theoretical framework of~\cite{Nicolai:2000sc,deWit:2002vt}.
Gaugings are defined by a constant embedding tensor that transforms in a particular representation
of the global symmetry group ${\rm E}_{d(d)}$. It is subject to a number of constraints and
entirely parametrizes the gauged Lagrangian. E.g.\ in the context of flux compactifications,
all possible higher-dimensional ($p$-form, geometrical, and non-geometrical)
flux components whose presence in the compactification
induces a deformation of the low-dimensional theory can be identified among the components of the embedding tensor.
Once the universal form of the gauged Lagrangian is known for generic embedding tensor, this 
reduces the construction of any particular example to a simple group-theoretical 
exercise.\footnote{Still, the 
explicit calculation of the various couplings from the closed formulas
may pose a considerable task.}
Moreover, since the embedding tensor combines the flux components
of various higher-dimensional origin into a single multiplet of the U-duality group ${\rm E}_{d(d)}$, 
this formulation allows to directly identify the transformation behavior of particular flux
components under the action of the duality groups. In particular, this allows to 
straightforwardly extend the analysis of the effective theories beyond the 
region in which the parameters have a simple perturbative or geometric interpretation.

Gaugings of two-dimensional supergravity ($d=9$) have not been studied systematically so far.
Yet, this case is particularly interesting, as the global symmetry algebra of the ungauged maximal theory
is the infinite-dimensional $\hat{\mathfrak{g}}=\mathfrak{e}_{9(9)}$, the affine extension of 
the exceptional algebra ${\mathfrak{g}}=\mathfrak{e}_{8(8)}$, and the resulting structures are extremely rich.
The realization of the affine symmetry on the physical fields requires the introduction of an infinite tower
of dual scalar fields, defined on-shell by a set of first order differential equations.
Consequently, these symmetries act nonlinearly, nonlocally and are 
symmetries of the equations of motion only.
As a generic feature of two-dimensional gravity theories, the
infinite-dimensional global symmetry algebra is a 
manifestation of the underlying integrable structure of the 
theory~\cite{Geroch:1970nt,Belinsky:1971nt,Maison:1978es,Korotkin:1997fi,Nicolai:1998gi}.
In view of the above discussion one may expect that the various parameters characterizing the different
higher-dimensional compactifications join into a single infinite-dimensional multiplet of the affine algebra 
which accordingly parametrizes the generic gauged Lagrangian in two dimensions. 
We confirm this picture in the present paper. The corresponding multiplet is
the basic representation of~$\mathfrak{e}_{9(9)}$.
\smallskip

Apart from its intriguing mathematical structure, there are two features of two-dimensional
supergravity which render the construction of gaugings somewhat more subtle than in higher dimensions.
First, the overwhelming part of the affine symmetries present in the two-dimensional ungauged
theory, is of the hidden type and in particular on-shell. Only the zero-modes $\mathfrak{g}$ of the 
affine algebra $\hat{\mathfrak{g}}$ are realized as target-space isometries of the two-dimensional scalar sigma-model
and thus as off-shell symmetries of the Lagrangian.
In contrast, the action of all higher modes of the algebra is nonlinear, nonlocal and on-shell as described above.
Gauging such symmetries is a nontrivial task.
Second, in two dimensions there are no propagating vector fields that could be naturally used to 
gauge these symmetries. 

It turns out that both these problems have a very natural common solution:
introducing a set of vector fields that couple with a particular topological 
term in the Lagrangian allows to gauge arbitrary subgroups of the affine symmetry group.
The resulting gauge groups generically include former on-shell symmetries and thus extend beyond
the target-space isometries of the ungauged Lagrangian.
The construction in fact is reminiscent of the four-dimensional case where global symmetries 
that are only on-shell realized can be gauged upon simultaneous introduction
of magnetic vector and two-form tensor fields which couple with topological 
terms~\cite{deWit:2005ub,dWST4}.

The structure emerging in two dimensions is the following. 
In addition to the original physical fields, the Lagrangian of the gauged theory
carries vector fields $A_{\mu}^{{\cal M}}$ in a highest weight representation of $\hat{\mathfrak{g}}$.
In addition, a finite subset of the tower of dual scalar fields enters the Lagrangian, with their defining
first-order equations arising as genuine equations of motion.
The gauging is completely characterized by a constant embedding tensor  $\Theta_{{\cal M}}$ in the 
conjugate vector representation and subject to a quadratic consistency constraint.
The local gauge algebra is a generically infinite-dimensional subalgebra of $\hat{\mathfrak{g}}$.
The result is a Lagrangian that features scalars and vector fields in infinite-dimensional 
representations of the affine~$\hat{\mathfrak{g}}$.
However, for every particular choice of the embedding tensor
only a finite subset of these fields enters the Lagrangian
and only a finite-dimensional part of the gauge algebra is
realized at the level of the Lagrangian 
(with its infinite-dimensional part exclusively acting on dual scalar fields 
that do not show up in the Lagrangian).
We illustrate these structures with several examples for the maximal ($N=16$) theory 
for which the symmetry algebra is $\mathfrak{e}_{9(9)}$
and vector fields and embedding tensor transform in the basic representation
and its conjugate, respectively.

In addition to the standard minimal couplings within covariant derivatives and the
new topological term, the gauging induces a scalar potential whose explicit form is usually 
determined by supersymmetry.
It is specific to two dimensions that in absence of such a potential, 
the gauging merely induces a reformulation of the original theory. 
I.e.\ the field equations imply vanishing field strengths, such that the only nontrivial effect of the 
newly introduced vector fields is due to global obstructions. In absence of such,
the theory reduces to the original one. On the other hand, integrating out the vector fields in this case
leads to an equivalent (T-dual) formulation of the original theory in terms of a different set of scalar fields.
This procedure is well-known from the study of non-Abelian 
T-duality~\cite{Buscher:1987sk,Hull:1989jk,delaOssa:1992vc}, however the results here
go beyond the standard expressions, as the gaugings generically include non-target-space isometries.
In contrast, in presence of a scalar potential, as is standard in supersymmetric theories,
the gaugings constitute genuine deformations of the original theory.

It is worth to stress that although the construction we present in this paper is worked out
for a very particular class of two-dimensional models --- the coset space sigma-models
coupled to dilaton gravity as the typical class of models obtained by dimensional reduction
of supergravity theories --- it is by far not limited to this class.
The entire construction extends straightforwardly to the gauging of hidden
symmetries in arbitrary two-dimensional integrable field theories.
\smallskip

The paper is organized as follows. In section~2, we give a brief review of the ungauged
two-dimensional supergravity theories and their global symmetry structure. In particular,
we give closed formulas for the action of the affine symmetry $\hat{\mathfrak{g}}$ 
on the physical fields.
In section~3, we proceed to gauge subalgebras of the affine global symmetry by
introducing vector fields in a highest weight representation of $\hat{\mathfrak{g}}$
and coupling them with a particular topological term. 
We present the full bosonic Lagrangian which is entirely parametrized in terms of
an embedding tensor
 transforming under $\hat{\mathfrak{g}}$ in the conjugate vector field representation
 and subject to a single quadratic constraint.
In section~4, we discuss various ways of gauge fixing part of the local symmetries 
by eliminating some of the redundant fields from the Lagrangian.
In particular, we show that in absence of a scalar potential
the presented construction leads to an equivalent (T-dual) version of the ungauged theory
whereas a scalar potential leads to genuinely inequivalent deformations of the original theory.
Finally, in section~5 we study various examples of gaugings of the maximal ($N=16$)
two-dimensional supergravity.
Among the infinitely many components of the embedding tensor, 
we identify several solutions to the quadratic constraint and discuss their 
higher-dimensional origin.
The various gradings of $\mathfrak{e}_{9(9)}$ provide a systematic scheme
for this analysis.

%++++++++++++++++++++++++++++++++++++++++++++++++++++++++++++++++++++++++++++++++++++++++++++++++++
\section{Ungauged theory and affine symmetry algebra}
%++++++++++++++++++++++++++++++++++++++++++++++++++++++++++++++++++++++++++++++++++++++++++++++++++

The class of theories we are going to study in this paper are two-dimensional ${\rm G}/{\rm K}$ coset space
sigma models coupled to dilaton gravity. These models arise from dimensional reduction of 
higher-dimensional gravities: pure Einstein gravity in four space-time dimensions gives rise to the
coset space ${\rm SL}(2)/{\rm SO}(2)$ while e.g.\ the bosonic sector of eleven-dimensional supergravity
leads to the particular coset space ${\rm E}_{8(8)}/{\rm SO}(16)$.
In this chapter we briefly review the Lagrangian for these theories, their integrability structure,
and as a consequence of the latter
the realization of the infinite-dimensional on-shell symmetry $\hat{\mathfrak{g}}$,
cf.~\cite{Breitenlohner:1986um,Nicolai:1991tt,Nicolai:1996pd} for detailed accounts.

%%%%%%%%%%%%%%%%%%%%%%%%%%%%%%%%%%%%%%%%%%%%%%%%%%%%%%%%%%%%%%%%%%%%%%%%%%%%%%%%%%
\subsection{Lagrangian}
%%%%%%%%%%%%%%%%%%%%%%%%%%%%%%%%%%%%%%%%%%%%%%%%%%%%%%%%%%%%%%%%%%%%%%%%%%%%%%%%%%

To define the Lagrangian of the theory we employ the decomposition
$\mathfrak{g}=\mathfrak{k} \oplus \mathfrak{p}$ of the 
Lie algebra $\mathfrak{g}={\rm Lie}\,{\rm G}$ into its compact part~$\mathfrak{k}$
and the orthogonal non-compact complement~$\mathfrak{p}$. 
For the theories under consideration 
this is a symmetric space decomposition, i.e.\
the commutators are of the form
\begin{align}
   [ \mathfrak{k} , \mathfrak{k} ] &= \mathfrak{k} \; , &
   [ \mathfrak{k} , \mathfrak{p} ] &= \mathfrak{p} \; , &
   [ \mathfrak{p} , \mathfrak{p} ] &= \mathfrak{k} \; .
\end{align}
We denote by $t_{\alpha}$ the generators of ${\mathfrak g}$ and
indicate by subscripts the projection onto the subspaces
$\mathfrak{k}$ and $\mathfrak{p}$, i.e.\ for
$\Lambda \in \mathfrak{g}$ it is
\begin{align}
   \Lambda &= \Lambda^{\alpha}t_{\alpha}=\Lambda_\mathfrak{k} + \Lambda_\mathfrak{p} \; , \qquad \qquad
   \Lambda_\mathfrak{k} \in \mathfrak{k} \; , \qquad
   \Lambda_\mathfrak{p} \in \mathfrak{p} \; .
   \label{DefHK}
\end{align}
In addition, it is useful to introduce the following involution on algebra elements
\begin{align}
   \Lambda^{\#} &=  \Lambda_\mathfrak{k}  - \Lambda_\mathfrak{p} \; .
\end{align}
The (${\rm dim}\,{\rm G}-{\rm dim}\,{\rm K}$)
bosonic degrees of freedom of the theory are described by a group element ${\cal V}$ of ${\rm G}$
which transforms under global ${\rm G}$
transformations from the left and local ${\rm K}$ transformations from the right, i.e.\
the theory is invariant under
\begin{align}
  {\cal V} \, & \rightarrow \, g \, {\cal V} \, k(x)^{-1} \; , \qquad \qquad
  g \in{\rm G} \; , \quad k(x) \in {\rm K} \; .
  \label{coset}
\end{align}
It is sometimes convenient to fix the local ${\rm K}$ freedom by restricting to a particular set of
representatives ${\cal V}$ of the coset ${\rm G}/{\rm K}$, on which the global ${\rm G}$ then acts as
\begin{align}
  {\cal V} \, & \rightarrow \, g \, {\cal V} \, k_{g}(x)^{-1} \;,
  \label{cosetNL}
\end{align}
where $k_{g}(x)\in {\rm K}$ depends on $g$ in order to preserve the class of representatives.
This defines the {\em nonlinear realization} of ${\rm G}$ on the coset space ${\rm G}/{\rm K}$.

The ${\rm G}$-invariant scalar currents are defined by
\begin{align}
   {\cal V}^{-1} \partial_{\mu} {\cal V} &~=~ Q_{\mu} + P_{\mu}  \; , \qquad \qquad
   Q_{\mu} \in \mathfrak{k} \; , \qquad
   P_{\mu} \in \mathfrak{p} \; .
   \label{Defcurrents}
\end{align}
The current $Q_{\mua}$ is a
composite connection for the local ${\rm K}$ gauge invariance, i.e.\ it appears in covariant derivatives
of all quantities that transform under ${\rm K}$, in particular
\begin{align}
   D_{\mua} P_{\mub} &= \partial_{\mua} P_{\mub} + [Q_{\mua},P_{\mub}] \; .
\end{align}
The integrability conditions for \eqref{Defcurrents} are then given by 
\begin{align}
   D_{[{\mua}} P_{{\mub}]} &= 0 \; , &
   Q_{{\mua}{\mub}} \, &\equiv \, 2 \partial_{[{\mua}} Q_{{\mub}]} + [Q_{\mua},Q_{\mub}] = - [P_{\mua},P_{\mub}] \; .
   \label{IntConPQ}   
\end{align}
The two-dimensional Lagrangian takes the form
\bea
   {\cal L} &=& \partial_\mua {\sigma}\,  \partial^\mua \rho
\, - \, \ft12 \rho \, \tr(P_\mua P^\mua) \; .
   \label{EffLconfD2}        
\eea
In addition to the scalar current $P_\mua$ it contains the dilaton field $\rho$
and the conformal factor $\sigma$. 
The latter originates from the two-dimensional metric which has been brought into conformal gauge  
$g_{\mu\nu}=e^{2\sigma}\,\eta_{\mu\nu}$,
such that space-time indices $\mu$ in (\ref{EffLconfD2}) are contracted 
with the flat Minkowski metric $\eta_{\mu\nu}$.
The only remnant of two-dimensional gravity is the first term descending from the 
two-dimensional (dilaton coupled) Einstein-Hilbert term $\rho R$ in conformal gauge.
The Lagrangian~(\ref{EffLconfD2}) is manifestly invariant under the symmetry (\ref{coset}).
It is straightforward to derive the equations of motion
which take the form\footnote{Our space-time conventions are 
$\eta_{\mu\nu}={\rm diag}(+,-)$, $\epsilon_{{01}}=-\epsilon^{01}=1$; 
i.e.\ $\eta_{\pm\mp}=1$, $\epsilon_{\pm\mp}=\mp$. }
\begin{align}
   \partial_{+}\partial_{-} \rho &= 0 \; , &
   \partial_{+}\partial_{-} {\sigma} + \ft 1 2 \, \tr(P_+ P_{-}) &= 0 \; , &
   D_+(\rho P_{-}) +D_-(\rho P_{+})&= 0 \;,
   \label{UngaugedEOMd2}
\end{align}
where we have introduced light-cone coordinates $x^\pm=(x^0 \pm x^1)/\sqrt{2}$.
In addition, the theory comes with two first order (Virasoro) constraints
\begin{align}
  \partial_\pm \rho \, \partial_\pm \sigma-  \ft12\rho \, \tr(P_\pm P_\pm) &= 0 \;,
   \label{ConfConst2}
\end{align}
which might equally be obtained from the Lagrangian before the fixing of conformal gauge.
It is straightforward to check that these first order constraints are compatible as a consequence
of the equations of motion for $\rho$ and $P_{\pm}$ and moreover imply the second order
equation for the conformal factor $\sigma$.

%%%%%%%%%%%%%%%%%%%%%%%%%%%%%%%%%%%%%%%%%%%%%%%%%%%%%%%%%%%%%%%%%%%%%%%%%%%%%%%%%%
\subsection{Global symmetry and dual potentials}
%%%%%%%%%%%%%%%%%%%%%%%%%%%%%%%%%%%%%%%%%%%%%%%%%%%%%%%%%%%%%%%%%%%%%%%%%%%%%%%%%%

It is well known --- starting from the work of Geroch 
on dimensionally reduced Einstein gravity~\cite{Geroch:1970nt,Kinnersley:1977ph,Cosgrove:1980fx}
--- 
that the global symmetry algebra of the coset space sigma model \eqref{EffLconfD2}
is not only the algebra of target-space isometries  $\mathfrak{g}$, but half of 
its affine extension~$\hat{\mathfrak{g}}$~\cite{Julia:1981wc}.
We denote the generators of $\mathfrak{g}$ by $t_\alpha$ and
those of $\hat{\mathfrak{g}}$ by $T_{\alpha,m}$, $m\in\mathbb{Z}$.
The latter close into the algebra~\cite{Kac:1990gs}
\bea
{}\Big[\;T_{\alpha,m}\;,\;T_{\beta,n}\:\Big] &=& f_{\alpha\beta}{}^{\gamma}\,T_{\gamma,m+n}+
m\,\delta_{m+n}\,\eta_{\alpha\beta}\,{ K}
\;,
\label{affine}
\eea
where $f_{\alpha\beta}{}^{\gamma}$ and $\eta_{\alpha\beta}={\rm tr}(t_{\alpha}t_{\beta})$
are the structure constants and the Cartan-Killing form of ${\mathfrak g}$, respectively, 
and $K$ denotes the central extension of the affine algebra. In addition to $T_{\alpha,m}$ and $K$
we will find the Witt-Virasoro generator $L_1$ to be crucial for the construction of this paper.
It obeys
\bea
[\,L_1,T_{\alpha,m}\,]&=& -m\,T_{\alpha,m+1} \;.
\label{affine2}
\eea
The central extension ${K}$ commutes with both $T_{\alpha,m}$ and $L_1$.
We denote by $\mathfrak{G}\supset\hat{\mathfrak{g}}$ 
the algebra spanned by $\{T_{\alpha,m}\,,K,L_{1}\}$.

To define the action of $\mathfrak{G}$ on the fields
${\cal V}$, $\rho$ and $\sigma$ that enter the Lagrangian \eqref{EffLconfD2} we need to introduce
an infinite hierarchy of dual potentials.
These are additional scalar fields that are defined as nonlocal functions
of ${\cal V}$ (and $\rho$), but whose definition is only consistent if one invokes the
equations of motion. Therefore $\mathfrak{G}$ is only realized
as an on-shell symmetry on~\eqref{UngaugedEOMd2}.

To start with, the dilaton $\rho$ is a free field, such that it gives rise to the definition
\bea
 \partial_\mu \tilde\rho &=& -\epsilon_{\mu\nu} \partial^{\nu} \rho \;,
 \qquad
 \Longleftrightarrow
 \qquad 
 \partial_\pm \tilde\rho ~=~ \pm \partial_\pm \rho \; ,
   \label{DualDilaton}
\eea
of its dual $\tilde \rho$.
Obviously, the dual of $\tilde \rho$ gives back~$\rho$.
More interesting are the nonlinear equations of motion for ${\cal V}$
that can be rewritten as a conservation law $\partial^\mua I_\mua =0$ for the current 
$I_\mua = \rho\,{\cal V} P_\mua {\cal V}^{-1}$.
This allows the definition of the first dual potential $Y_{1}$
\bea
   \partial_\pm Y_1 &=&  \mp I_{\pm}~=~\mp \rho {\cal V} P_\pm {\cal V}^{-1} \;,
   \label{DefY1}
\eea
which is ${\mathfrak g}$ valued and according to (\ref{coset}) 
transforms in the adjoint representation of the global ${\rm G}$.
Integrability of these equations is ensured by $\partial^\mua I_\mua =0$. 
From the point of view of higher-dimensional supergravity theories,
equations~\eqref{DefY1} constitute nothing but a particular case of the general on-shell 
duality between $p$ forms and $D-p-2$ forms ($D=2$, $p=0$).
In two dimensions however, these equations are just 
the starting point for an infinite hierarchy of dual potentials
of which the next members $Y_{2}$, $Y_{3}$ are defined by
\begin{align}
   \partial_\pm Y_2 &=  \left( \pm \rho \tilde \rho + \ft 1 2 \rho^2 \right) {\cal V} P_\pm {\cal V}^{-1} 
                       + \ft 1 2 [Y_1,\partial_\pm Y_1]  \; ,
\nonumber \\
   \partial_\pm Y_3 &=  \left( \mp \ft 1 2 \rho^3 \mp \rho \tilde \rho^2 - \rho^2 \tilde \rho \right) {\cal V} P_\pm {\cal V}^{-1}
                       + [Y_1,\partial_\pm Y_2] - \ft 1 6 [Y_1,[Y_1,\partial_\pm Y_1]]] \; .
   \label{DefY23}       
\end{align}
Again, integrability of these equations is guaranteed by the field equations
$\partial^\mua I_\mua =0$ and the defining equation~\eqref{DefY1} of the lower dual potentials.
A convenient way to encode the definition of all dual potentials (and the action of 
the affine symmetry) is the {\em linear system}~\cite{Belinsky:1971nt,Maison:1978es}
which we will describe in the next subsection. 
In order  make the symmetry structure more transparent we will
restrict the discussion in the present subsection to the 
lowest few dual potentials and to the action of the lowest
few affine symmetry generators $T_{\alpha,m}$.

We identify the zero-modes $T_{\alpha,0}$ with the generators $t_\alpha$
of the off-shell symmetry~$\mathfrak{g}$.
These zero-mode symmetries do not mix the
original scalars and the dual potentials of different levels,
i.e.\ ${\cal V}$ transforms according to \eqref{coset} and
all the $Y_m$ ($m>0$) transform in the adjoint representation. The 
fields $\rho$, $\tilde \rho$, and $\sigma$
are left invariant by $T_{\alpha,0}$.

The dual potentials $\tilde \rho$, $Y_m$ are defined by~\eqref{DualDilaton}--\eqref{DefY23}
only up to constant shifts $\tilde \rho \mapsto \tilde \rho + \lambda$, $Y_m \mapsto Y_m + \Lambda_m$.
The generators in $\mathfrak{G}$ corresponding to these shift symmetries are 
$L_1$ and $T_{\alpha,m}$ ($m>0$), i.e.
\begin{align}
   \delta_{(1)} \,\tilde \rho &= 1 \; , &
   \delta_{\alpha,m} \,Y^\beta_n &= 
   \left\{ 
   \begin{array}{ll}
   \delta_\alpha^\beta & m=n\\
   0 & m>n
   \end{array}\right.
    \; ,    
   \label{ShiftSym}
\end{align}
where $\delta_{(1)}$ and $\delta_{\alpha,m}$ denote the action of $L_1$
and $T_{\alpha,m}$, respectively,
and $Y_{m}=Y^\alpha_m t_{\alpha}$\,.
Since the definition of the dual potentials also involves
$\tilde \rho$ and lower dual potentials, it follows that $L_1$ and
$T_{\alpha,m}$ also act nontrivially on the higher dual potentials $Y_n$ ($m<n$), e.g. 
\begin{align}
   \delta_{(1)}\, Y_2 &= - Y_1 \; ,  &
   \delta_{(1)}\, Y_3 &= - 2 Y_2 \; ,  \nonumber \\
   \Lambda^\alpha \, \delta_{\alpha,1}\, Y_2 &= \ft 1 2 [\Lambda,Y_1] \; , & &
   \mbox{etc.}
   \label{ShiftSym2}
\end{align}
None of the shift symmetries $L_1$ and $T_{\alpha,m}$ ($m>0$) acts on 
the physical fields ${\cal V}$, $\rho$ or $\sigma$. So far we have thus not introduced
any new physical symmetry. 
The crucial point about the symmetry structure of the model is
the existence of another infinite family of symmetry generators $T_{\alpha,m}$ ($m<0$).
Their action on the physical fields is expressed in terms of the dual potentials
and thus nonlinear and nonlocal in terms of the original fields.
For the lowest generators, this action is given by
\begin{align}
   \Lambda^{\alpha} \delta_{\alpha,-1}\, {\cal V} &=
   [\Lambda,Y_{1}] {\cal V} 
      - \tilde \rho \; {\cal V} [{\cal V}^{-1} \Lambda {\cal V}]_{\mathfrak{p}}
    \; , \nonumber \\ 
   \Lambda^{\alpha}\delta_{\alpha,-2}\, {\cal V} &=
   \left\{ [\Lambda,Y_{2}] + \ft 1 2 [[\Lambda,Y_{1}],Y_{1}] -
               \tilde \rho [\Lambda,Y_{1}] \right\} {\cal V}
      + \left(\ft 1 2 \rho^2 + \tilde \rho^2\right) \, {\cal V} [{\cal V}^{-1} \Lambda {\cal V}]_{\mathfrak{p}}
    \;.
    \nonumber \\
   \label{NonLocalSym}
\end{align}
The field $\rho$ is left invariant while the action on the dual potentials $Y_{m}$
and on the conformal factor $\sigma$ follows 
from~\eqref{ConfConst2}, \eqref{DefY1}. We find for example
\begin{align}
   \Lambda^{\alpha}\delta_{\alpha,-1}\, \sigma &= \tr (\Lambda Y_1)
  \; , 
    \nonumber \\
       \Lambda^{\alpha}\delta_{\alpha,-1}\, Y_{1} &=
      [\Lambda,Y_2] + \ft 1 2 [[\Lambda,Y_{1}],Y_{1}]
         + \ft 1 2 \rho^2  {\cal V} [{\cal V}^{-1} \Lambda {\cal V}]_{\mathfrak{p}} {\cal V}^{-1}  \; ,
   \qquad
   \mbox{etc.} 
\end{align}
One can easily check that the symmetries defined in \eqref{ShiftSym} and \eqref{NonLocalSym}
indeed close according to the algebra \eqref{affine}. 
In particular, it follows that the central extension $K$
acts exclusively on the conformal factor~\cite{Julia:1981wc}:
\bea
   \delta_{(0)}\, \sigma &= \, - 1 \;.
   \label{centralK}
\eea

In order to define all dual potentials $Y_m$ ($m>0$) and describe the action of all
symmetry generators $T_{\alpha,m}$ in closed form we will in the following 
introduce the linear system~\cite{Belinsky:1971nt,Maison:1978es} 
showing the classical integrability of the theory.

%%%%%%%%%%%%%%%%%%%%%%%%%%%%%%%%%%%%%%%%%%%%%%%%%%%%%%%%%%%%%%%%%%%%%%%%%%%%%%%%%%
\subsection{The linear system}
%%%%%%%%%%%%%%%%%%%%%%%%%%%%%%%%%%%%%%%%%%%%%%%%%%%%%%%%%%%%%%%%%%%%%%%%%%%%%%%%%%

A compact way to encode the infinite family of dual potentials and the 
action of the full symmetry algebra~$\hat{\mathfrak{g}}$
is the definition of a one-parameter family of group-valued matrices 
$\hV(\gamma)$
according to the linear system~\cite{Belinsky:1971nt,Maison:1978es,Breitenlohner:1986um}
\bea
   \hV^{-1} \partial_\mu \hV \, &=& \hat{J}_{\mu}\;,
   \qquad\mbox{with}\quad
   \hat{J}_{\mu}~=~
   Q_{\mu}+ \frac{1+ \gamma^{2}} {1- \gamma^{2}} \,P_\mu
  +\frac{2\gamma}{1- \gamma^{2}} \,\epsilon_{\mu\nu}\, P^{\nu} 
\;,
   \label{LinearSystem}
\eea
where $\gamma$ is a scalar function 
\begin{align}
   \gamma &= 
   \frac 1 \rho \left( w + \tilde \rho - \sqrt{(w+\tilde \rho)^2 - \rho^2} \right)  \;,
    \label{DefGamma}
\end{align}
of the {\em constant spectral parameter} $w$ which labels the family.
As $\gamma$ is a double-valued function of $w$ we will in the following restrict to
the branch $|\gamma|<1$, i.e.\ in particular
\bea
\gamma &=& 
\ft 1 2 \rho \,w^{-1} - \ft 1 2 \rho \tilde \rho \,w^{-2} 
+ \ft 1 8 \left( \rho^3 + 4 \rho \tilde \rho^2 \right)  w^{-3}+\dots \;,
\label{gammaser}
\eea
around $w=\infty$.

It is straightforward to verify that 
the compatibility of~(\ref{LinearSystem}) is equivalent to~\eqref{IntConPQ} 
and the equations of motion~(\ref{UngaugedEOMd2}):
\bea
2\partial_{[\mu}\hat{J}_{\nu]}+[\hat{J}_{\mu},\hat{J}_{\nu}]&=&
Q_{\mu\nu}+[P_{\mu},P_{\nu}]
+  \frac{1+ \gamma^{2}} {1- \gamma^{2}} \,2D_{[\mu}P_{\nu]}
  -\epsilon_{\mu\nu}\, \frac{2\gamma}{1- \gamma^{2}} \,\rho^{-1}\,
  D^{\sigma}(\rho P_{\sigma}) \;.
  \nonumber\\
  \label{dJ}
\eea
Expanding $\hV$ around $w=\infty$
\bea
   \hV &=& \ldots\; e^{w^{-4}\,Y_4 } e^{w^{-3}\,Y_3} e^{w^{-2}\,Y_2} e^{w^{-1}\,Y_1} \,{\cal V} \; ,
   \label{ExpandHV}
\eea
defines the infinite series of dual potentials $Y_{n}$.
In particular, the expansion of~\eqref{LinearSystem} around $w=\infty$
reproduces~\eqref{DefY1},~\eqref{DefY23}. 
For later use we also give the 
linear system in light-cone coordinates
 \bea
  \hV^{-1} D_\pm \hV \, 
   &=&
    \frac{1\mp \gamma} {1\pm \gamma} \, P_\pm 
   \;.
   \label{LinearSystem2}
\eea

Using the matrix~$\hV$, the action of the symmetry algebra~$\mathfrak{G}$ can
be expressed in closed form. To this end, we parametrize the loop 
algebra of ${\mathfrak{g}}$
by a spectral parameter~$w$ and identify the generators $T_{\alpha,m}$
with $w^{-m} t_\alpha$.
Elements $\Lambda=\Lambda^{\alpha,m} T_{\alpha,m}$ of
$\hat{\mathfrak{g}}$ are represented by $\mathfrak{g}$-valued functions 
$\Lambda(w)=\Lambda^{\alpha,m} w^{-m} t_{\alpha}$,
meromorphic in the spectral parameter plane.
In terms of $\Lambda(w)$, the action on the physical fields ${\cal V}$, $\sigma$ 
can be given in closed form as
\bea
{\cal V}^{-1}\,\delta_\Lambda {\cal V} &=&
     \left\langle
     \frac{2 \gamma(w)} {\rho\,(1-\gamma^{2}(w))} \, \hat \Lambda_{{\mathfrak p}}(w) \; \right\rangle_{w}\;,
 \nonumber \\ [1.5ex]
\delta_\Lambda \, \sigma &=&
     \, - \, \tr \, \Big\langle \Lambda(w) \, \partial_w  \hV(w) \, \hV^{-1}(w) \Big\rangle_{w} \;.
    \label{E9action}   
\eea
Here we have defined the dressed parameter\footnote{
For notational simplicity we use here and in the following
the notation $\hV(w)\equiv\hV(\gamma(w))$, even though by definition globally $\hV$
is a function of $\gamma$ and thus on the double covering of the complex $w$-plane. 
We will however be mainly interested
in its local expansion around $w=\infty$ on the sheet~\eqref{gammaser}.
}
\bea
\hat \Lambda(w)&=&  \hV^{-1}(w) \Lambda(w) \hV(w)~=~
\hat \Lambda_{{\mathfrak k}}(w)+\hat\Lambda_{{\mathfrak p}}(w)\;,
\eea
with the split according to~\eqref{DefHK}. 
In addition, we have introduced the notation
\bea
   \langle f(w) \rangle_w  &\equiv&
   \oint_\ell \frac {dw} {2 \pi i} \,f(w) = - \text{Res}_{w=\infty} \,f(w) \; ,
\label{path}
\eea
for an arbitrary function $f(w)$ of the spectral parameter $w$.
The path $\ell$ is chosen such that only the residual at $w=\infty$ is picked up. 
For definiteness we will treat the functions $f(w)=\sum_{m=-\infty}^{\infty} f_m w^{m}$
in these expressions as formal power series 
with almost all $\{f_{m}|m>0\}$ equal to zero. 
Some useful relations for calculating with these objects are collected in appendix~\ref{App:algebra}.

It is straightforward to check that the transformations~(\ref{E9action})
leave the equations of motion invariant.
Since the solution $\hV(w)$
of the linear system~(\ref{LinearSystem}) explicitly enters the 
transformation, this is in general not a symmetry of the Lagrangian
but only an on-shell symmetry of the equations of motion~(\ref{UngaugedEOMd2}). This will
be of importance later on. Moreover, it is straightforward to check,
that the algebra of transformations~(\ref{E9action})
closes according to~\eqref{affine}. Relation~\eqref{rel2}
is crucial to verify the action~\eqref{centralK} of the central extension.

The group-theoretical structure of the symmetry~(\ref{E9action})
becomes more transparent if we consider its extension to $\hV(w)$
and thereby to the full tower of dual potentials~\cite{Korotkin:1997fi}:
\bea
     \hV^{-1}\, \delta_\Lambda  \hV(w) &=& \hat \Lambda (w) 
                               - \left\langle \frac 1 {v-w} \left( \hat \Lambda_{\mathfrak k}(v)
                               + \frac{\gamma(v)\,(1-\gamma^2(w))}{\gamma(w)\,(1-\gamma^2(v))} \;
                                                   \hat \Lambda_{\mathfrak p}(v) \right) \right\rangle_{v} \;,
\label{E9actionVH}
\eea
in the above notation. This action may be rewritten as
\bea
\delta_\Lambda  \hV(w) &=& 
\Lambda (w)\,\hV(w) - \hV(w)\,\Upsilon(\gamma(w))\;,
\label{E9actionVH2}
\\[1ex]
&&
{\rm with}\;\;\;
\Upsilon(\gamma(w))\equiv
{\textstyle\left\langle \frac 1 {v-w} \left( \hat \Lambda_{\mathfrak k}(v)
 + \frac{\gamma(v)\,(1-\gamma^2(w))}{\gamma(w)\,(1-\gamma^2(v))} \;
 \hat \Lambda_{\mathfrak p}(v) \right) \right\rangle_{v}} \;,
\nonumber
\eea
and thus takes the form of an infinite-dimensional analogue of
the nonlinear realization~(\ref{cosetNL}),
in which the left action of $\Lambda(w)$ parametrizing $\hat{\mathfrak g}$ is 
accompanied by a right action of $\Upsilon(\gamma)\in {\mathfrak k}(\hat{\mathfrak g})$ 
in order to preserve a particular class of coset representatives.
The algebra ${\mathfrak k}(\hat{\mathfrak g})$ 
is the infinite-dimensional analogue of ${\mathfrak k}$ in~(\ref{cosetNL}),
i.e.\ the maximal compact subalgebra of $\hat{\mathfrak{g}}$,
and is defined as the algebra 
of ${\mathfrak g}$-valued functions $k(\gamma)$, satisfying~\cite{Breitenlohner:1986um}\footnote{
Note that ${\mathfrak k}(\hat{\mathfrak g})\not=\hat{\mathfrak k}$.}
\begin{align}
   k^{\#}(\gamma) &= k(1/\gamma) \; .
\end{align}
We shall see in the following that 
the particular set of coset representatives starring in~(\ref{E9actionVH2}) 
are the functions $\hV(\gamma(w))$ regular
around $w=\infty$ in accordance with the expansion~(\ref{ExpandHV}).

For illustration, let us
evaluate equation~(\ref{E9actionVH2}) for 
the particular transformation 
$\Lambda(w)=w^{-m}\,\Lambda$,\, $\Lambda\in{\mathfrak g}$, $m\in\mathbb{Z}$\,.
Expanding both sides around $w=\infty$, it 
follows directly from~(\ref{rel1}) that for positive values of $m$,
$\Upsilon(\gamma)$ vanishes, such that the transformation merely amounts to a shift of
the dual potentials $Y_{n}$ in the expansion~(\ref{ExpandHV}); for $m=1, 2$ this 
reproduces~\eqref{ShiftSym},~\eqref{ShiftSym2}.
These transformations do not act on the physical
fields present in the Lagrangian~(\ref{EffLconfD2}).
For a transformation with negative $m$ on the other hand the second term
in~(\ref{E9actionVH2}) no longer vanishes but precisely restores
the regularity of $\hV$ at $w=\infty$ that has been destroyed by the first 
term~\cite{Nicolai:2004nv}.
These transformations describe the nonlinear and nonlocal on-shell symmetries
on the physical fields and the dual potentials which leave the equations of motion
and the linear system~(\ref{LinearSystem2}) invariant. 
They are commonly referred to as hidden symmetries, for $m=-1$
one recovers~\eqref{NonLocalSym}.
Finally, for $m=0$ one recovers the action~(\ref{coset}) of the finite 
algebra ${\mathfrak g}$
acting as an off-shell symmetry on all the fields.
Here, the local ${\rm K}$ freedom in~(\ref{coset}) has been fixed such that 
$[{\cal V}^{-1}\delta{\cal V}]_{{\mathfrak k}}=0$.

To summarize, the negative modes $T_{\alpha,m}$, $m<0$ act as nonlocal on-shell symmetries
whereas the positive modes $T_{\alpha,m}$, $m>0$ act as shift symmetries on the dual potentials.
Only the zero-modes $T_{\alpha,0}$ are realized as off-shell symmetries on the physical fields 
of the Lagrangian~(\ref{EffLconfD2}).

In addition to the affine symmetry algebra $\hat{\mathfrak g}$
described above, a Witt-Virasoro algebra can be realized on the fields~\cite{Julia:1996nu}
which essentially acts as conformal transformation on the inverse spectral parameter $y=1/w$.
{}From these generators we will in the following only need 
\bea
L_1 = - y^2 \partial_y = \partial_w \;,
\eea
which acts only on the dual dilaton $\tilde\rho$ and
the dual potentials $Y_n$ according to equations \eqref{ShiftSym},~\eqref{ShiftSym2}
\bea
   \delta_{(1)} \tilde \rho = 1 
   \quad \Longrightarrow\quad 
   \delta_{(1)} \hV =  \, \partial_w  \, \hV \; . 
   \label{L1action}
\eea

The pair $K$ and $L_1$  which extends the loop algebra of ${\mathfrak{g}}$
to $\mathfrak{G}$ turns out to be crucial 
for our construction of the gauged theory in section~\ref{sec:gauging}. 
The distinguished role of $L_{1}$ in this construction 
--- as opposed to all the other Virasoro generators
that can be realized following~\cite{Julia:1996nu} ---
stems from its action on the dual dilaton~(\ref{ShiftSym}). 
The gaugings we are mainly interested in
will carry a scalar potential whose presence in particular 
deforms the free field equation~(\ref{UngaugedEOMd2})
of $\rho$ by some source terms $\Box\rho=Q$. 
The only way to maintain a meaningful version of the dual 
dilaton equation~(\ref{DualDilaton}) in this case is
by gauging its shift symmetry $\partial_{\mu}\rho=-\epsilon_{\mu\nu}(\partial^{\nu}
 - {\cal B}^{\nu}\,\delta_{(1)})\, \tilde \rho$ while imposing 
 $\partial_{[\mu}{\cal B}_{\nu]}=-\epsilon_{\mu\nu}\,Q$.
We shall see that this indeed 
appears very natural in the subsequent construction.

In the following we will parametrize a general algebra 
element of $\mathfrak{G}\equiv\langle T_{\alpha,m}\,, K, L_{1}\rangle$ 
with a collective label $\cAa\in\{(\alpha,m),(1),(0)\}$ 
for the generators of $\mathfrak{G}$ as
\bea
   \Lambda &=& \Lambda^\cAa \,T_\cAa ~=~ 
   \Lambda^{\alpha,m}\, T_{\alpha,m} \, 
   +  \, \Lambda^{(1)} \, L_{1} \, + \, \Lambda^{(0)} \, K 
   ~\equiv~ 
   \Lambda(w)  
   +   \Lambda^{(1)}  L_{1}  +  \Lambda^{(0)} \, K\;,
\nonumber\\
   \label{SymParamD2}
\eea
with $\Lambda(w)\equiv \Lambda^{\alpha,m}w^{-m}\,t_{\alpha}$.
The commutator between two such algebra elements takes the form
\bea
{}\leftb\Lambda,\Sigma\,\rightb &=&
[\Lambda(w),\Sigma(w)]+
\Lambda^{(1)}\partial\Sigma(w)-
\Sigma^{(1)}\partial\Lambda(w)+
K\,\Big\langle \Lambda(w)\,\partial\Sigma(w)\Big\rangle_{\!w}
\;,
\label{fullcomm}
\eea
where we use the notation $\leftb,\rightb$ in order to distinguish the 
general algebra commutator from the simple matrix commutators $[\,,]$.

Let us finally mention, that the symmetry algebra 
$\mathfrak{G}$
is equipped with an invariant inner product 
$(T_\cAa,T_\cAb)=\eta_{\cAa\cAb}$, given by
\bea
(T_{\alpha,m}\,,T_{\beta,n})&=&\eta_{\alpha\beta}\,\delta_{m+n-1}\;,\qquad
(L_{1},K)=-1\;.
\label{invbil}
\eea
Note that this invariant form differs from the standard one by the shift of $-1$
in the $L_{0}$ grading. This is precisely consistent with the use of $L_{1}$ rather
than $L_{0}$ in the pairing with the central extension $K$.

%++++++++++++++++++++++++++++++++++++++++++++++++++++++++++++++++++++++++++++++++++++++++++++++++++
\subsection{Structure of the duality equations}
%++++++++++++++++++++++++++++++++++++++++++++++++++++++++++++++++++++++++++++++++++++++++++++++++++
\label{structure}

For the following it turns out the be important to analyze in more detail
the structure of the duality equations \eqref{DualDilaton} and \eqref{LinearSystem}
which have been used to define the dual fields $\tilde\rho$ and $\hV$.
Let us for the moment consider these dual fields as a priori independent fields
and the duality equations as their first order equations of motion 
relating them to the physical fields $\rho$ and ${\cal V}$.
In particular, we may define the $\mathfrak{G}$-valued current $Z_{\mu}$ as
\bea
 { Z}_\mua &=& { Z}^{\cAa}_\mua\,T_{\cAa}~=~
   { Z}_{\mua}(w) \, + \, { Z}^{(1)}_\mua\, L_{1}\;,
    \label{defZ}\\[2ex]
   { Z}^{(1)}_\mua &\equiv& 
- {\partial}_{\mua} \tilde \rho     - \, \epsilon_{\mua\mub}\partial^\mub \rho \; ,
  \nonumber\\
   { Z}_\mua(w) &\equiv& 
  \hV \,\Big[ 
  -  \hV^{-1} {\partial}_\mua \hV  +Q_{\mua}
 + \frac{ 1+\gamma^2 } {1 - \gamma^2} \, {P}_\mua
 + \frac {2 \gamma} {1-\gamma^2}\,  \epsilon_{\mua\mub}{ P}^\mub 
\Big] \hV^{-1}
\, - \, \partial_w \hV \, \hV^{-1} \, { Z}^{(1)}_\mua   \;,
\nonumber
\eea
which is a particular combination of the duality equations, i.e.\
on-shell we have $Z_{\mu}=0$. 
Under a generic symmetry transformation $\Lambda\in{\mathfrak G}$
the constituents of $Z_{\mu}$ transform according to
(\ref{E9action}), (\ref{E9actionVH}), and (\ref{L1action})
and some lengthy computation shows that altogether $Z_{\mu}$
transforms as
\bea
   \delta_{\Lambda} Z_{\pm} &=&  
   \leftb\Lambda,Z_{\pm}\rightb
     - \hV\;\Big \langle \frac 1 {v-w} \,\hV^{-1}\, \leftb\Lambda, {Z}_{\pm}\rightb\,\hV
     \Big\rangle_{{\mathfrak{k}},v}\hV^{-1}
     \nonumber\\[.7em]
     &&{}\qquad
            - \frac{1\mp\gamma} {1\pm\gamma}\;
            \hV \,\Big\langle\frac 1 {v-w} \, \frac{1\pm\gamma} {1\mp\gamma}\,
                \hV^{-1}\, \leftb\Lambda, {Z}_{\pm}\rightb\,\hV
     \Big\rangle_{{\mathfrak{p}},v}\hV^{-1}
     \;,
     \label{deltaZ}
\eea
in light-cone coordinates.
In order not to overburden the notation here,
all spectral parameter dependent functions within the brackets $\langle\cdot\rangle_{v}$
depend on the parameter $v$ which is integrated over, 
whereas all functions outside depend on the spectral parameter~$w$.
In slight abuse of notation, the commutators 
$\leftb,\rightb$ represent the full $\mathfrak{G}$ commutator~\eqref{fullcomm}
however without the 
central term $K$.\footnote{Inclusion of this term would presumably require the extension of $Z_{\mu}$
by a $K$-valued term proportional to the Virasoro constraints~\eqref{ConfConst2}.
This is in accordance with the generalized linear system proposed in~\cite{Bernard:1997et}.
For the purpose of this paper however this would complicate things unnecessarily.}
In particular, \eqref{deltaZ} shows that
$Z_{\mu}$ transforms homogeneously under $\Lambda$
--- consistent with the fact that $Z_{\mu}$ vanishes on-shell.
This current will play an important role in the following.

%++++++++++++++++++++++++++++++++++++++++++++++++++++++++++++++++++++++++++++++++++++++++++++++++++
\section{Gauging subgroups of the affine symmetry}
%++++++++++++++++++++++++++++++++++++++++++++++++++++++++++++++++++++++++++++++++++++++++++++++++++
\label{sec:gauging}

In the previous section
we have reviewed how the equations of motion of the ungauged two-dimensional theory
are invariant under an infinite algebra ${\mathfrak G}$ of symmetry transformations.
The symmetry action on the physical fields~(\ref{E9action}) is defined in terms of 
the matrix $\hV$ which in turn is defined as a solution of the linear system \eqref{LinearSystem}.
As a result, the global symmetry is nonlinearly and nonlocally realized on the physical fields.

We will now attempt to gauge part of the global symmetry~(\ref{E9action}),
i.e.\ turn a subalgebra of $\mathfrak{G}$ into a local symmetry of the theory. 
This is rather straightforward for subalgebras of 
$\mathfrak{g}=\langle T_{\alpha,0}\rangle\subset\mathfrak{G}$,
as $\mathfrak{g}$ is the off-shell symmetry algebra of the Lagrangian.
In fact, since $\mathfrak{g}$ is already the off-shell symmetry of the three-dimensional
ancestor of the theory, the corresponding gaugings are simply obtained
by dimensional reduction of the three-dimensional gauged 
supergravities~\cite{Nicolai:2000sc,deWit:2003ja}.
The gauging of generic subalgebras of $\mathfrak{G}$ is much more intricate, 
as their action explicitly contains the matrix $\hV$ which is defined only on-shell
as a nonlocal functional of the physical fields. 
This is the main subject of this paper.
The problem is analogous to the one faced in four dimensions when trying
to gauge arbitrary subgroups of the scalar isometry group -- not restricting to
triangular symplectic embeddings -- which has been solved only recently~\cite{deWit:2005ub,dWST4}.
We will follow a similar approach here.

As a key point in the construction we will introduce the
dual scalars $\tilde{\rho}$ and $\hV$ as independent fields on the Lagrangian level.
The duality equations~\eqref{defZ} relating them to the original fields will 
naturally emerge as first order equations of motion.
Specifically, the field equations obtained by varying the Lagrangian with respect to 
the newly introduced gauge fields of the theory turn out to be proportional to the
current $Z_{\mu}$ introduced in section~\ref{structure} which 
combines the duality equations.

\subsection{Gauge fields and embedding tensor}

In order to construct the gauged theory, 
we make use of the formalism of the embedding tensor,
introduced to describe the gaugings of supergravity in higher 
dimensions~\cite{Nicolai:2000sc,deWit:2002vt}.
Its main feature is the description of the possible gaugings in a formulation 
manifestly covariant under the global symmetry $\mathfrak{G}$ of the ungauged theory.
As a first step we need to introduce vector fields in order
to realize the covariant derivatives corresponding to the local symmetry. 
In contrast to higher dimensions where the vector fields come in some well-defined representation
of the global symmetry group of the ungauged theory, in two dimensions
these fields do not represent propagating degrees of freedom
and are absent in the ungauged theory.\footnote{Also in three dimensions
it is most convenient to start from a formulation of the ungauged theory
in which no vector fields are present~\cite{Nicolai:2000sc,deWit:2003ja}. 
In contrast to the present case, however,
the vector fields in three dimensions are dual to the scalar fields and thus
naturally come in the adjoint representation of the scalar isometry group.}
We will hence start by introducing a set of vector fields $A_{\mu}^{\cal M}$
transforming in some a priori undetermined representation 
(labeled by indices~{\footnotesize ${\cal M}$}) of the algebra~$\mathfrak{G}$.

An arbitrary gauging then is described by an {\em embedding tensor} $\Theta_{{\cal M}}{}^{{\cal A}}$
that defines the generators
\bea
X_{{\cal M}}&\equiv& \Theta_{{\cal M}}{}^{{\cal A}}\,T_{{\cal A}}
\;,
\label{genX}
\eea
of the subalgebra of $\mathfrak{G}$ which is promoted to a local symmetry by introducing covariant derivatives
\bea
{\cal D}_\mua &=
\partial_\mua - g \, {\cal A}_\mua^{{\cal M}} \, 
\Theta_{{\cal M}}{}^{{\cal A}}\,T_{{\cal A}} \;,
\label{covD}
\eea
with a gauge coupling constant $g$.\footnote{The coupling constant $g$
always comes homogeneous with the embedding tensor and could simply be 
absorbed by rescaling $\Theta_{{\cal M}}{}^{{\cal A}}$. We will keep it explicitly 
to have the deformation more transparent.} 
The way $\Theta_{{\cal M}}{}^{{\cal A}}$
appears within these derivatives shows that  under ${\mathfrak{G}}$ it naturally transforms
in the tensor product of two infinite-dimensional representations. Gauge invariance immediately
imposes the quadratic constraint (or embedding equation)
\bea
f_{{\cal BC}}{}^{{\cal A}} \, \Theta_{{\cal M}}{}^{{\cal B}}\,\Theta_{{\cal N}}{}^{{\cal C}} +
    T_{{\cal B}, {\cal N}}{}^{{\cal P}}\,
    \Theta_{{\cal M}}{}^{{\cal B}} \, \Theta_{{\cal P}}{}^{{\cal A}} &=&0
\;,
\label{ConstraintQuadratic}
\eea
on $\Theta_{{\cal M}}{}^{{\cal A}}$,
where $f_{{\cal BC}}{}^{{\cal A}}$ denote the structure constants of the algebra~(\ref{affine}),
(\ref{affine2}), and $T_{{\cal B}, {\cal N}}{}^{{\cal P}}$ are the generators of ${\mathfrak G}$
in the representation of the vector fields.
Equivalently, this constraint takes the form
\bea
{}[X_{{\cal M}},X_{{\cal N}}] &=& -X_{{\cal MN}}{}^{{\cal K}}\,X_{{\cal K}} \;,
\label{algebra}
\eea
with ``structure constants''\footnote{
We have put quotation marks here because according to this definition 
the constants $X_{{\cal MN}}{}^{{\cal K}}$
are not antisymmetric in the first two indices, but only after further multiplication
with a generator $X_{{\cal K}}$. 
Manifest antisymmetrization on the other hand defines objects $X_{[{\cal MN}]}{}^{{\cal K}}$ that do no longer
satisfy the Jacobi identities. Analogous structures arise in higher-dimensional 
gauged supergravity theories~\cite{de Wit:2005hv}.}
$X_{{\cal MN}}{}^{{\cal K}}=\Theta_{{\cal M}}{}^{{\cal A}}\,T_{{\cal A,N}}{}^{{\cal K}}$.
We will impose further constraints on $\Theta_{{\cal M}}{}^{{\cal A}}$ in the sequel.

It will sometimes be convenient to expand
the covariant derivatives~(\ref{covD}) according to~(\ref{SymParamD2}) as
\begin{align}
{\cal D}_\mua &= 
   \partial_\mua 
- g \, {\cal A}^\alpha_\mua(w) \, t_\alpha 
- g \, {\cal A}^{(1)}_\mua \, L_{1} 
- g \, {\cal A}^{(0)}_\mua \, K \; ,
\label{covD1}
\end{align}
with the projected vector fields
\bea
   {\cal A}^{(1)}_\mua = \Theta_{{\cal M}}{}^{{(1)}} \, A_\mua^{{\cal M}}   \;,
   \qquad
   {\cal A}^{(0)}_\mua = \Theta_{{\cal M}}{}^{{(0)}} \, A_\mua^{{\cal M}}   \;,
     \qquad
   {\cal A}^\alpha_\mua(w) = 
   \sum_{m=-\infty}^{m=\infty} w^{-m} \,\Theta_{{\cal M}}{}^{{\alpha,m}}
   \, A_\mua^{{\cal M}}  \; .
   \label{projV}
\eea
While the appearance of the infinite sums (over $m$ and over ${\cal M}$) in the definition of 
${\cal A}^\alpha_\mua(w)$ (and thus the appearance of an infinite number of vector fields)
looks potentially worrisome, we will eventually impose constraints on $\Theta_{{\cal M}}{}^{{\alpha,m}}$
such that only a finite subset of vector fields $A_\mua^{{\cal M}}$ enters the Lagrangian.

Explicitly, the action of the covariant derivative on the various scalars reads\footnote{
Comparing~(\ref{covD3}) to~(\ref{Defcurrents}) one notices that 
${\cal Q}_\mua\equiv [{\cal V}^{-1} {\cal D}_\mua {\cal V}]_{{\mathfrak{k}}}={Q}_\mua$
does not depend on the coupling constant $g$.
This is due to our particular ${\rm SO}(16)$ gauge
choice in equation \eqref{E9action}.}
\begin{align}
    {\cal D}_\mua \tilde \rho &= \partial_\mua \tilde \rho - g\,{\cal A}^{(1)}_\mua\;,
        \nonumber \\[1ex]
    {\cal D}_\mua \sigma &= \partial_\mua \sigma + g\,{\cal A}^{(0)}_\mua + 
    g\,\tr \, \Big\langle {\cal A}_\mua(w)\, \partial_w \hV(w) \hV^{-1}(w) \Big\rangle_{w}\;, 
        \nonumber \\[1ex]
    {\cal V}^{-1} {\cal D}_\mua {\cal V} & = {\cal V}^{-1} \partial_\mua {\cal V} - g\,\Big\langle
     \frac{2 \gamma(w)} {\rho\,(1-\gamma^{2}(w))} \, \hat {\cal A}_\mua(w)_{\mathfrak{p}} \; \Big\rangle_{w}
        \; = \;  {\cal P}_\mua + {\cal Q}_\mua     \; ,
        \nonumber \\[2ex]
   \hV^{-1} {\cal D}_\mua \hV(w) &= \hV^{-1} \partial_\mua \hV(w) - g\,{\cal A}^{(1)}_\mua \hV^{-1} \partial_w \hV(w)
                             - g\,\hat {\cal A}_\mua (w)  \nonumber \\[.5ex] & \qquad \qquad 
                             + g\,\left\langle \frac 1 {v-w} \left( [\hat {\cal A}_\mua(v)]_{\mathfrak k}  
                               + \frac{\gamma(v)\,(1-\gamma^2(w))}{\gamma(w)\,(1-\gamma^2(v))}\,
                                                   [\hat {\cal A}_\mua(v)]_{\mathfrak p} \right) \right\rangle_{v} \; ,
\label{covD3}
\end{align}
with $\hat {\cal A}_\mua(w) = \hV^{-1}(w) {\cal A}_\mua(w) \hV(w)$.

\subsection{The Lagrangian}

As a first step towards introducing the local symmetry on the level of the Lagrangian,
we consider the covariantized version of~(\ref{EffLconfD2})
\bea
   {\cal L}_{\rm{kin}} &=& 
   \partial^\mua\! \rho \, {\cal D}_\mua \sigma - \ft 1 2 \, \rho \, \tr ( {\cal P}_\mua {\cal P}^\mua ) \; ,
\label{Lkin}
\eea
with covariant derivatives according to~(\ref{covD3}). Obviously,~(\ref{Lkin})
cannot be the full answer since the equations of motion for the newly introduced 
vector fields will pose unwanted (and in general inconsistent) first order relations among
the scalar fields. Likewise, according to \eqref{covD3} 
the ${\cal P}_{\mu}$ now carry the dual potentials $\tilde{\rho}$ and $\hV$ which 
are to be considered as independent fields. Variation with respect to these fields then
gives rise to even stranger constraints.

Remarkably, all these problems can be cured by adding to the Lagrangian
what we will refer to as a {\em topological term}\footnote{We call this term topological as 
after relaxing conformal gauge it does not depend on the two-dimensional metric.}
\bea
   {\cal L}_{\rm{top}} &=&
    -g\,\epsilon^{\mua\mub} \;\Big\{ 
    \tr\, \Big\langle \hat{\cal A}_{\mua}\, 
  \Big(\hV^{-1}(\partial_{\mub}\hV-\partial_{w}\hV\, \partial_\mub \tilde \rho )
      - Q_{\nu}
      - \frac{1+\gamma^{2}}{1-\gamma^{2}}\,P_{\mub}
      \Big) \Big\rangle_{\!w}\,
      -  {\cal A}^{(0)}_{\mua}\,\partial_\mub \tilde \rho\; \Big\}
     \nonumber \\ &&
                - \ft 1 2 \, g^2 \,\epsilon^{\mua\mub}\, {\cal A}^{(0)}_\mua {\cal A}^{(1)}_\mub   
                - \ft 1 2 \, g^2 \,\epsilon^{\mua\mub}\,
\tr\, \Big\langle \!\Big \langle \;
\frac 1 {v-w}\, [\hat {\cal A}_\mua(w)]_{\mathfrak k} \, [\hat {\cal A}_\mub(v)]_{\mathfrak k}
\, \Big \rangle_{\!v} \:\Big \rangle_{\!w} 
\label{Ltop} \\[1ex] &&
- \ft 1 2 \, g^2 \,\epsilon^{\mua\mub}\,
\tr\, \Big\langle \!\Big \langle \;
 \frac {(\gamma(v)-\gamma(w))^2 + (1-\gamma(v)\gamma(w))^2} {(v-w) (1-\gamma^{2}(v)) (1-\gamma^{2}(w))}\:
[\hat {\cal A}_\mua(w)]_{\mathfrak p} \, [\hat {\cal A}_\mub(v)]_{\mathfrak p}
                      \,  \Big \rangle_{\!v} \:\Big \rangle_{\!w}   \; ,
                        \nonumber
\eea
which is made such that the vector field equations of motion precisely yield (a projection of)
the covariantized version of the duality equations~(\ref{DualDilaton}), (\ref{LinearSystem}).
Explicitly, the variation of the Lagrangian ${\cal L}_{0}={\cal L}_{\rm{kin}}+{\cal L}_{\rm{top}}$ 
with respect to the vector fields reads
\bea
   \delta {\cal L}_{0} &=& 
            -g \, 
\eta_{\cAa\cAb}\,\Theta_{{\cal M}}{}^{{\cAa}}\,\epsilon^{\mua\mub}
{\cal Z}_{\mua}^{\cAb}\,\delta A_\mub^{{\cal M}} \; ,
   \label{VarVec}    
\eea
where ${\cal Z}_{\mu}$ is the properly covariantized version of the
${\mathfrak{G}}$-valued current defined in~\eqref{defZ} above.
It contains the 
covariantized versions of the duality equations \eqref{DualDilaton} and \eqref{LinearSystem}
that render $\tilde\rho$ dual to $\rho$ and $\hV$ dual to ${\cal V}$, respectively. 
As vector field equations in the gauged theory 
we thus find a $\Theta$-projection of ${\cal Z}_\mua=0$\,:
\bea
g\,\Theta_{{\cal M}}{}^{{\cAa}}\,\eta_{{\cAa\cAb}}\,{\cal Z}_{\mua}^{\cAb}&=&0\;.
\label{projZ}
\eea
In the limit $g\rightarrow0$ back to the ungauged theory
these equations consistently decouple.
 
The fact that the higher order $g$ terms of~\eqref{projZ}
can be consistently integrated to
the variation~\eqref{VarVec} is nontrivial
and puts quite severe constraints on the construction.
Namely, it requires the following constraint 
\bea
   \tr\, \Big\langle {\cal A}_\mua(w) \, \delta {\cal A}_\mub(w) \Big \rangle_w 
          - {\cal A}^{(1)}_\mua \, \delta {\cal A}^{(0)}_\mub - {\cal A}^{(0)}_\mua \, \delta {\cal A}^{(1)}_\mub  &=&  0 \;,
\eea
on the variation with respect to the projected vector fields.
Fortunately, this condition translates directly into the $\mathfrak{G}$ covariant constraint
\bea
\Theta_{{\cal M}}{}^{{\cal A}}\,\Theta_{{\cal N}}{}^{{\cal B}}\,\eta_{{\cal AB}}
&=& 0 \;,
\label{Quadratic2}
\eea
for the embedding tensor~$\Theta_{{\cal M}}{}^{{\cal A}}$.
For consistency, this constraint must thus be imposed together with
the quadratic constraint~(\ref{ConstraintQuadratic})
ensuring gauge invariance.
As in higher-dimensional gaugings~\cite{deWit:2002vt}, we expect that the 
latter constraint~\eqref{Quadratic2}
should eventually be a consequence of~(\ref{ConstraintQuadratic}). 
This is one motivation for the ansatz
\bea
\Theta_{{\cal M}}{}^{{\cal A}}&=& 
T_{\cAb,{\cal M}}{}^{{\cal N}}\,\eta^{\cAa\cAb}\,\Theta_{{\cal N}}
\;,
\label{ConstraintLinear}
\eea
for the embedding tensor parametrized by a single conjugate vector $\Theta_{{\cal M}}$.
In terms of $\mathfrak{G}$ representations this means that $\Theta_{{\cal M}}{}^{{\cal A}}$ 
does not take arbitrary values in the tensor product of the coadjoint
and the conjugate vector field representation, but only in the 
conjugate vector field representation contained in this tensor product.
This is the analogue of the linear representation constraint that is
typically imposed on the embedding tensor in higher 
dimensions~\cite{Nicolai:2000sc,deWit:2002vt}.
Indeed, it is straightforward to verify that the ansatz~(\ref{ConstraintLinear}) 
reduces the quadratic constraints~(\ref{ConstraintQuadratic}) and (\ref{Quadratic2}) to the same 
constraint for $\Theta_{{\cal M}}$:
\bea
\eta^{\cAa\cAb}\,T_{\cAa,{\cal M}}{}^{{\cal P}}\,T_{\cAb,{\cal N}}{}^{{\cal Q}}\,
\Theta_{{\cal P}}\Theta_{{\cal Q}}
&=& 0 \;.
\label{Quadratic3}
\eea
Further support for the ansatz~(\ref{ConstraintLinear})
comes from the fact that all the examples of gauged theories
in two dimensions (presently known to us) turn out to 
be described by an embedding tensor of this particular form. 
In particular, in all examples
originating by dimensional reduction from a higher-dimensional gauged theory,
the constraint (\ref{ConstraintLinear}) is a consequence of the corresponding 
linear constraint in higher dimensions.
We will come back to this in section~5.
This shows that~(\ref{ConstraintLinear})
describes an important class of if not all the two-dimensional gaugings.

It is useful to give the projected vector fields~(\ref{covD1}) 
using~(\ref{ConstraintLinear})
\bea
   {\cal A}^{(1)}_\mua &=&  
   - T_{(0),\,{\cal M}}{}^{{\cal N}} \, A_\mua^{{\cal N}}  \, \Theta_{{\cal M}} \; , 
 \qquad  
 {\cal A}^{(0)}_\mua \,=\,  - \, T_{(1),\,{\cal M}}{}^{{\cal N}} \, A_\mua^{{\cal M}}  \, \Theta_{{\cal N}}  \; ,
     \nonumber \\[0.1cm]
   {\cal A}^\alpha_\mua(w) &=& 
   \sum_{m=-\infty}^{m=\infty} 
        w^{-m} \, \eta^{\alpha\beta} \,(T_{\beta,(1-m)})_{{\cal M}}{}^{{\cal N}} \, 
        A_\mua^{{\cal M}}  \, \Theta_{{\cal N}}~\equiv~  {\cal A}^{\alpha,m}_{\mu}\,w^{-m}\; .
   \label{DefABC} 
\eea
This further suggests that the vector fields $A_\mua^{{\cal M}}$ transform 
in some irreducible {\em highest weight} representation of ${\mathfrak{G}}$. 
Namely, in that case there is for any given ${\cal M}$ an integer $M$ such that
\bea
(T_{\beta,m})_{{\cal N}}{}^{{\cal M}} =0\;,  \qquad \mbox{for all\;\;} m>M
\;.
\eea
Formula~(\ref{DefABC}) then shows that for every gauging defined by an
embedding tensor $\Theta_{{\cal M}}$ with only finitely many non-vanishing entries, 
the projected vector fields ${\cal A}^\alpha_\mua(w)$ carry only finitely many 
positive powers of $w$. 
As a consequence, 
only finitely many of the $A_\mua^{{\cal M}}$ enter
the Lagrangian~(\ref{Lkin}), (\ref{Ltop}), which is certainly indispensable
for a meaningful action.

Moreover, it follows from~\eqref{ExpandHV} that the terms $\partial_w \hV \hV^{-1}$
and $\hV {\cal Z}_\mua(w) \hV^{-1}$ have expansions in $1/w$ 
starting with $w^{-2}$ and $w^{-1}$, respectively. From the variation~\eqref{VarVec} 
we thus find that the positive mode vector fields~${\cal A}_\mua^{\alpha,m}$, 
$m>0$, do not enter the Lagrangian at all. I.e.\ a gauging of the shift symmetries of the dual
potentials is not visible in the Lagrangian. 
{}From the Lagrangian itself this fact is not obvious since
the quadratic constraint was used to derive \eqref{VarVec}. 
Only a truncation of the full gauge group is thus manifest in the Lagrangian.
We will see this realized in explicit examples in section~\ref{Sec:maximal}.

In the rest of this section, we will show that every embedding tensor of the form~(\ref{ConstraintLinear})
with $\Theta_{{\cal M}}$ satisfying~(\ref{Quadratic3}) defines a gauge invariant Lagrangian.

\subsection{The quadratic constraint}
\label{sec:qc}

Let us pause for a moment and reconsider the present construction.
We have constructed the gauged Lagrangian~\eqref{Lkin}, \eqref{Ltop}
by covariantizing the ungauged theory and adding a topological term 
such that variation with respect to the new gauge fields yields the
scalar duality equations. The gauging is entirely
parametrized in terms of the embedding tensor $\Theta_{{\cal M}}$.
At first sight the formalism of the embedding tensor  
may seem unnecessarily heavy in two dimensions. As the new gauge fields 
enter the Lagrangian only in the contracted form 
$A_{\mu}^{{\cal A}}\equiv A_{\mu}^{{\cal M}}\,\Theta_{{\cal M}}{}^{{\cal A}}$,
could we not have started right away from a set of vector fields $A_{\mu}^{{\cal A}}$ in
the adjoint representation 
rather than introducing $A_{\mu}^{{\cal M}}$
in some yet undetermined representation, and 
$\Theta_{{\cal M}}{}^{{\cal A}}$ separately?
The answer is no. Consistency of the construction essentially depends on
the quadratic constraint~(\ref{Quadratic3}) on the embedding tensor
which in particular implies that not all components of
the projected $A_{\mu}^{{\cal A}}$ are independent. 
This is most conveniently taken care of 
by explicitly introducing~$\Theta_{{\cal M}}{}^{{\cal A}}$.

Before proceeding with the proof of gauge invariance of the Lagrangian,
we will in this subsection closer analyze this quadratic constraint imposed 
on the embedding tensor. It can be skipped on first reading.
We have shown above that the linear ansatz~(\ref{ConstraintLinear}) for $\Theta_{{\cal M}}{}^{{\cal A}}$
reduces the quadratic constraints~(\ref{ConstraintQuadratic}) and~(\ref{Quadratic2}) to the same 
constraint 
\bea
\eta^{\cAa\cAb}\,T_{\cAa,{\cal M}}{}^{{\cal P}}\,T_{\cAb,{\cal N}}{}^{{\cal Q}}\,
\Theta_{{\cal P}}\Theta_{{\cal Q}}
&=& 0 \;,
\label{Quadratic3A}
\eea
for the tensor $\Theta_{{\cal M}}$. This exhibits an interesting 
representation structure underlying the quadratic constraint. 
Formally, the constraint~(\ref{Quadratic3A}) lives in the 
twofold symmetric tensor product of the conjugate vector field representation.
In particular, if $\Theta_{{\cal M}}$ transforms in a level $k$ highest weight representation,
the constraint transforms in an (infinite) sum of level $2k$ highest weight representations.
As we are dealing with infinite-dimensional representations, 
these are most conveniently described in terms of the associated characters.
Let us denote by  $\chi_{\Theta}$ the character of the conjugate vector field representation,
and by $\chi_{{{i}}}$ the characters associated with the different level $2k$
representations ${\cal R}_{i}$ of $\widehat{\mathfrak{g}}$. They are extended to representations 
of the Virasoro algebra by means of the standard Sugawara construction.
In terms of these characters, the decomposition of the product $\Theta_{{\cal M}}\Theta_{{\cal N}}$ 
takes the form
\bea
\chi_{\Theta} \otimes_{{\rm sym}} \chi_{\Theta}
&=&
\sum_{i} \chi^{{\rm vir}}_{i}\cdot\chi_{{{i}}}
\;,
\label{prodvec}
\eea
where the sum is running over the level $2k$ representations of $\widehat{\mathfrak{g}}$
and the coefficients $\chi^{{\rm vir}}_{i}$ encoding the multiplicities of these representations
carry representations of the Virasoro algebra
associated with the coset model~\cite{Goddard:1984vk} 
\bea
      \frac {\widehat{\mathfrak{g}}_k \oplus \widehat{\mathfrak{g}}_k}{ \widehat{\mathfrak{g}}_{2k} } 
\;.
\label{cosetCFT}
\eea
For simplicity, we restrict to simply-laced Lie algebras $\mathfrak{g}$ in the following.
With the central charge of the Virasoro algebra on $\widehat{\mathfrak{g}}_k$ given by
   $c_k=k \, \text{dim}(\mathfrak{g})/({k+g^\vee}) $\,
in terms of the dual Coxeter number  $g^\vee$ of $\mathfrak{g}$,
the coset CFT has central charge
\bea
\frac{2k^{2} \, \text{dim}(\mathfrak{g})} {(k+g^\vee)(2k+g^\vee)} 
\;.
\label{CFTcentral}
\eea
The coset Virasoro generators acting on (\ref{prodvec}) are given by
\begin{align}
   L^{\text{coset}}_m &= L^{\widehat{\mathfrak{g}}_k \oplus \widehat{\mathfrak{g}}_k}_m 
   - L^{\widehat{\mathfrak{g}}_{2k}}_m \;,
\end{align}
in terms of the Virasoro generators induced by ${\widehat{\mathfrak{g}}_k \oplus \widehat{\mathfrak{g}}_k}$
and ${\widehat{\mathfrak{g}}_{2k}}$, respectively. A brief calculation reveals
that they take the explicit form
\bea
   (L^{\text{coset}}_m)_{\cMa\cMb}{}^{\cMc\cMd}
     &=& \frac{2} {k+g^\vee} \Big(  (L_m)_{(\cMa}{}^{(\cMc} \, \delta_{\cMb)}^{\cMd)} 
                                    - \sum_{n=0}^{\infty} \, \eta^{\alpha\beta} \, 
(T_{\alpha,m+n})_{(\cMa}{}^{(\cMc} \, (T_{\beta,-n})_{\cMb)}{}^{\cMd)} \Big) \;.
\nonumber
\eea
In particular, we thus obtain
\bea
   (L^{\text{coset}}_1)_{\cMa\cMb}{}^{\cMc\cMd}
     &=& - \, \frac{1} {k+g^\vee} \, \eta^{\cAa\cAb} \, T_{\cAa,\cMa}{}^{(\cMc} \, T_{\cAb,\cMb}{}^{\cMd)} \;,
\eea
which shows that the quadratic constraint (\ref{Quadratic3A}) can be rewritten
in strikingly compact form as
\begin{align}
   L^{\text{coset}}_1 ( \Theta \otimes \Theta) &= 0 \;.
\label{QuadraticL}
\end{align}
The quadratic constraint thus takes the form of a projector
on the product decomposition~(\ref{prodvec})
which acts on the multiplicities $\chi^{{\rm vir}}_{i}$.
Only those components within $\Theta$ whose products 
induce a quasi-primary state in the coset CFT~(\ref{cosetCFT})
give rise to a consistent gauging.
While this CFT formulation of the quadratic constraint is certainly very appealing
we do at present have no good interpretation for the appearance of this structure.
We will show explicitly in the next subsection
that~(\ref{QuadraticL}), alias~\eqref{Quadratic3A}, is a sufficient constraint for 
gauge invariance of the Lagrangian.

\subsection{Gauge invariance of the Lagrangian}

The Lagrangian~(\ref{Lkin}), (\ref{Ltop})
was determined above by requiring that variation with respect to the vector fields
yields a properly covariantized version of the scalar duality equations.
In particular, this uniquely fixes all higher order $g$ couplings.
In the rest of this section we will show that this Lagrangian 
is indeed invariant
under the local action of the generators~(\ref{genX})
\bea
\delta_{\Lambda} \, \tilde\rho &=& g\Lambda^{(1)}\;,
\nonumber\\[1ex]
\delta_{\Lambda} \, \sigma &=&
     \, - g\, \tr \, \Big\langle \Lambda(w) \, \partial_w  \hV(w) \, \hV^{-1}(w) \Big\rangle_{\!w} 
  -~ g\Lambda^{(0)}\;,
      \nonumber \\ [1.5ex]
{\cal V}^{-1}\,\delta_{\Lambda} {\cal V} &=&
     g\,\Big\langle
     \frac{2 \gamma(w)} {\rho\,(1-\gamma(w)^2)} \, \hat \Lambda_{{\mathfrak p}}(w) \; \Big\rangle_{\!w}\;,
 \nonumber \\ [1.5ex]
          \hV^{-1}\, \delta_{\Lambda}  \hV(w) &=& g\,\hat \Lambda (w) +g\,\Lambda^{(1)}\,{\hV}^{-1}\,\partial_{w} {\hV}
          \nonumber\\[.5ex]
&&                       {}-g \,\Big\langle \frac 1 {v-w} \Big( \hat \Lambda_{\mathfrak k}(v)
                               + \frac{\gamma(v)\,(1-\gamma^2(w))}{\gamma(w)\,(1-\gamma^2(v))} \;
                                                   \hat \Lambda_{\mathfrak p}(v) \Big) \Big\rangle_{\!v} 
\;,
\label{gaugescalar}
\eea
where 
\bea
\Lambda&=&
\Lambda^{{\cal M}}(x)\,\Theta_{{\cal M}}{}^{{\cal A}}\,T_{{\cal A}} 
~=~
\Lambda(w;x) +\Lambda^{(1)}(x)  \,L_{1}+\Lambda^{(0)}(x)\, K
\;,
\eea
now is a space-time dependent element of~$\mathfrak{G}$
induced by the gauge parameter $\Lambda^{{\cal M}}(x)$.
In addition, the 
action of the generators on the vector fields needs to be properly implemented.

To this end, we first compute the variation of 
${\cal L}_{0}={\cal L}_{{\rm kin}}+{\cal L}_{{\rm top}}$
under generic variation of vector and scalar fields. 
A somewhat tedious but beautiful computation shows that this variation may
be cast in the following compact form
\bea
   \delta {\cal L}_{0} &=&  
   -g \, T_{\cAa,\cMa}{}^\cMb \,  \Theta_\cMb \,
   \, \epsilon^{\mua\mub}{\cal Z}_\mua{}^\cAa\,(\Delta  A_\mub^{{\cal M}})
-  \partial_\mua \partial^\mua \rho\,\delta \sigma 
- \left( \widehat{\cal D}_\mua {\cal D}^\mua \sigma  
+ \ft 1 2 \tr {\cal P}_\mua {\cal P}^\mua \right)\,\delta \rho
     \nonumber \\[1ex]        
             &&{} + \tr \left(   \widehat{\cal D}_\mua (\rho {\cal P}^\mua) 
             \left[ {\cal V}^{-1} \delta {\cal V} \right]_\mathfrak{p}\right)
 - \ft 1 2 \, g\,\epsilon^{\mua\mub} 
\, T_{\cAa,\cMa}{}^\cMb \, \widehat{\cal F}_{\mua\mub}^\cMa \, \Theta_\cMb\;  \delta \hat\Sigma^\cAa 
 \;.
\label{VaryL3}
\eea
The quadratic constraint~(\ref{Quadratic3}) on $\Theta_{{\cal M}}$ is essential
in the derivation of this result.
In expressing the generic variation 
we have introduced the ``covariantized'' variations
\bea
   \Delta  A_\pm^\cMa &\equiv& \delta  A_\pm^\cMa 
                        + T_{\cAa,\cMb}{}^\cMa \,  A_\pm^\cMb\,\delta \hat\Sigma_\pm^\cAa  \;,
   \nonumber \\[1ex]
    \delta \hat\Sigma_\pm &\equiv& \hV\,\Big\{
        \hV^{-1} \delta \hV - [{\cal V}^{-1} \delta {\cal V}]_{\mathfrak{k}}
               - \frac{1\mp\gamma} {1\pm\gamma} \, [{\cal V}^{-1} \delta {\cal V}]_\mathfrak{p}
                          \Big\}\,\hV^{-1}  + (L_1 - \partial_w \hV \hV^{-1} )  \, 
                          (\delta \tilde \rho \mp\delta \rho)\; ,
   \nonumber \\[.6ex] 
     \delta \hat\Sigma&\equiv&  \ft12(\delta \hat\Sigma_++\delta \hat\Sigma_-)
     \;,
\label{varAS}
\eea   
and generalized field strength and covariant derivatives according to
\bea   
   \widehat{\cal F}_{\mua\mub}^\cMa &=& 2 \partial_{[\mua} { A}^\cMa_{\mub]} 
    ~- 2 \, T_{\cAa,\cMb}{}^\cMa \, {\cal Z}^\cAa_{[\mua} \, A_{\mub]}^\cMb
   ~+g X_{{\cal P Q}}{}^{{\cal M}}\,{ A}^{{\cal P}}_{[\mua} \,{ A}^{{\cal Q}}_{\mub]} 
   \;,
     \label{Diff3}    \\[1ex] 
   \widehat{\cal D}_\mua {\cal D}_\mub \sigma &=& \partial_\mua {\cal D}_\mub \sigma 
~- g\,A^\cMa_\mua \,  
    \left\{ \delta_\cMa^\cMb \, {\cal D}_\mub 
          + {\cal Z}_\mub^\cAc \, T_{\cAc,\cMa}{}^{\cMb}   \right\} 
\Theta_\cMb{}^{\cAa}  \, (T_{\cAa} \!\cdot\!\sigma)
   \; , \nonumber \\[1ex]
   \widehat{\cal D}_\mua {\cal P}_\mub &=& (\partial_\mua + \ad_{{\cal Q}_\mua}) {\cal P}_\mub 
              \nonumber \\[.6ex] 
              && {}
- g\,A^\cMa_\mua \,  
    \left\{ \delta_\cMa^\cMb \, ({\cal D}_\mub + \ad_{{\cal Q}_\mub})  
          + {\cal Z}_\mub^\cAc \, T_{\cAc,\cMa}{}^{\cMb}   \right\} 
   \Theta_\cMb{}^{\cAa} \, [{\cal V}^{-1} (T_\cAa\!\cdot\! {\cal V})]_{\mathfrak{p}}
    \;.
\nonumber
\eea
These expressions differ from the standard definitions of field strength and
covariant derivatives by the appearance of the current ${\cal Z}_{\mu}$
containing the duality equations of the ungauged theory. 
Recall that in the gauged theory only its $\Theta$-projection~(\ref{projZ})
is zero by the equations of motion.
Its natural appearance in~(\ref{Diff3})
motivates the introduction of generalized covariant derivatives~$\widehat{\cal D}$
\bea
\widehat{\cal D}_{\mu} &=& \partial_{\mu}+({\cal Z}_{\mu}^{{\cal A}}-
g\,A_{\mu}^{{\cal M}}\,\Theta_{{\cal M}}{}^{{\cal A}})\,T_{{\cal A}}
\;.
\label{fullcov}
\eea
Note that as ${\cal Z}_{\mu}$ contains only negative powers of $w$,
it only couples to shift symmetry generators in the covariant derivatives.
Thus, for all physical fields $\rho$, ${\cal V}$, there is no difference between the
full covariant derivative~$\widehat{\cal D}$ and (\ref{covD}) defined above.

In view of~\eqref{VaryL3}, \eqref{Diff3}, a natural ansatz for the transformation of the vector fields is
\bea
   \delta_{\Lambda} A_\mua^\cMa &=& 
   \widehat{\cal D}_\mua \Lambda^\cMa ~\equiv~
   {\cal D}_\mua \Lambda^\cMa 
         \, - \,  {\cal Z}_\mua^\cAa\,T_{\cAa,\cMb}{}^{\cMa} \, \Lambda^\cMb  \;  .
\label{gaugevector}
\eea
Indeed, the main result we establish in this section is the invariance
of the full Lagrangian ${\cal L}_{0}={\cal L}_{{\rm kin}}+{\cal L}_{{\rm top}}$
under the combined action
\eqref{gaugescalar}, \eqref{gaugevector}
of the local gauge algebra.

We now give a sketch of the proof. Computing 
the covariantized variations~\eqref{varAS} for the
gauge transformations~\eqref{gaugescalar} yields
\bea
   \delta_{\Lambda} \hat\Sigma &=& g\Lambda(w)-gk \Lambda^{{\cal M}}\Theta_{{\cal M}}\,L_1
 \nonumber\\  
&&{}   \!\!\!\!\!\!\!\!
-g\,    \hV(w)\,
        \Big\langle \frac 1 {v-w} \Big( \hat \Lambda_{\mathfrak k}(v)
 + \frac {(\gamma(v)-\gamma(w))^2 + (1-\gamma(v)\gamma(w))^2} { (1-\gamma^{2}(v)) (1-\gamma^{2}(w))} \;
 \hat \Lambda_{\mathfrak p}(v) \Big) \Big\rangle_{\!v}  \hV^{-1}(w)
  \; ,
  \nonumber
\eea   
and
\bea
\Delta_{\Lambda}\, A^{{\cal M}}_{\pm} &\!=\!&
\widehat{\cal D}_\pm \Lambda^\cMa 
          +( g\Lambda(w)-gk \Lambda^{{\cal M}}\Theta_{{\cal M}}\,L_1   )^\cAa\, T_{\cAa,\cMb}{}^\cMa \,  A_\pm^\cMb
          \label{extraA}\\[1ex]
          &&{}
-g\,    \Big(\hV\,
        \Big\langle \frac 1 {v-w} \hat \Lambda\Big\rangle_{{\mathfrak k},v}  \hV^{-1}
        +  \frac{1\pm\gamma}{1\mp\gamma}\,  \hV\,
        \Big\langle \frac 1 {v-w} \,\frac{1\mp\gamma}{1\pm\gamma}\,
 \hat \Lambda \Big\rangle_{\!{\mathfrak p},v}  \hV^{-1}\Big)^\cAa\; T_{\cAa,\cMb}{}^\cMa \,  A_\pm^\cMb
 \;.
 \nonumber
\eea
Again, we use the short-hand notation according to which
all spectral parameter dependent functions within the brackets $\langle\cdot\rangle_{v}$
depend on the parameter $v$ which is integrated over, 
whereas all functions outside depend on the spectral parameter~$w$.
Plugging all the variations into the Lagrangian, one obtains
after some lengthy computation and up to total derivatives
\bea
\delta_{\Lambda}\,{\cal L}_{0} &=& - \ft12\,g \,\Theta_{{\cal M}}{}^{{\cal A}}\,\eta_{{\cal A}{\cal B}}\,
\Lambda^{{\cal M}}\,\epsilon^{\mu\nu}\,
{\cal X}^{{\cal B}}_{\mu\nu}
\;,
\label{deltaL}
\eea
with
\bea
{\cal X}_{\mu\nu}&\equiv&
2{\cal D}_{[\mu}{\cal Z}_{\nu]}+
\leftb{\cal Z}_{\mu},{\cal Z}_{\nu}\rightb
+
2\widehat{\cal D}_{[\mu}{\cal J}_{\nu]}-\leftb{\cal J}_{\mu},{\cal J}_{\nu}\rightb
-g\,\widehat{\cal F}_{\mu\nu}{}^{{\cal M}}\,\Theta_{{\cal M}}{}^{{\cal A}}\,
(T_{{\cal A}}\cdot{\hV})\,\hV^{-1}
\;,\nonumber\\[2ex]
&&{}
{\cal J}_{\mu}
~\equiv~ \hV\,\Big\{\;{\cal Q}_\mu+
\frac{1+ \gamma^{2}} {1- \gamma^{2}}\,{\cal P}_\mu
  +\frac{2\gamma}{1- \gamma^{2}} \,\epsilon_{\mu\nu}\, {\cal P}^{\nu}\,\Big\}\;\hV^{-1} 
\;.
\label{J}
\eea
The calculation makes use of the covariantized version of~\eqref{dJ}
for $\hat{\cal J}_{\mu}=\hV^{-1}{\cal J}_{\mu}\hV$.
The subtle part in calculating~\eqref{deltaL} is the check that
the various terms arising from the different variations arrange into the
correct covariant derivatives, as the Lagrangian
and the variations have no manifest covariance. 
E.g.~the extra $A_{\mu}^{{\cal M}}$
contributions from~\eqref{extraA} are precisely the ones needed in order
to complete the correct covariant derivatives ${\cal D}_{\mu}$ on ${\cal Z}_{\nu}$ 
in ${\cal X}_{\mu\nu}$.
For this it is important to note that due to the extra contributions of order $g^{0}$ in~\eqref{gaugevector} 
the variation of ${\cal Z}_{\mu}$ changes with respect to the ungauged theory~\eqref{deltaZ}
to
\bea
   \delta_{\Lambda} {\cal Z}_{\pm} &=&  
   F(\Lambda,{\cal Z})
     - \hV\;\Big \langle \frac 1 {v-w} \,\hV^{-1}\, F(\Lambda,{\cal Z})\,\hV
     \Big\rangle_{{\mathfrak{k}},v}\hV^{-1}
     \nonumber\\[.7em]
     &&{}\qquad
            - \frac{1\mp\gamma} {1\pm\gamma}\;
            \hV \,\Big\langle\frac 1 {v-w} \, \frac{1\pm\gamma} {1\mp\gamma}\,
                \hV^{-1}\, F(\Lambda,{\cal Z})\,\hV
     \Big\rangle_{{\mathfrak{p}},v}\hV^{-1}
     \;,
     \nonumber\\[3ex]
     &&{}{\mbox{with}}\quad
     F(\Lambda,{\cal Z})^{{\cal A}}~\equiv~
     -g\,\Lambda^{M}\,({\cal Z}_{\mu}^{B}\,\Theta_{{\cal N}{\cal B}})
     \,T^{{\cal A}}{}_{{\cal M}}{}^{{\cal N}}\;,
     \label{deltaZgauged}
\eea
where indices {\small ${\cal A}$},  {\small ${\cal B}$}
are lowered and raised with $\eta_{{\cal A}{\cal B}}$ and its inverse.
Indeed, this is precisely consistent with the fact that in the gauged theory only the projection
${\cal Z}_{\mu}^{B}\,\Theta_{{\cal N}{\cal B}}$ vanishes on-shell as a
set of first order equations of motion for the dual potentials~\eqref{projZ} ---
accordingly, it must transform homogeneously under gauge transformations.

It remains to show that ${\cal X}_{\mu\nu}$ vanishes. In order to do so,
we first note that with the definition~\eqref{fullcov}
of generalized covariant derivatives $\widehat{\cal D}_{\mu}$, we find for the dual fields~$\tilde\rho$,~$\hV$
\bea
\widehat{\cal D}_{\mu}\tilde{\rho} &=& -\epsilon_{\mu\nu}\,\partial^{\nu}\rho
\;,
\nonumber\\[1ex]
\widehat{\cal D}_{\mu}\hV\,\hV^{-1} &=&{\cal J}_{\mu} 
   \;,
  \label{newcov}
\eea
with ${\cal J}_{\mu}$ from ~\eqref{J}, changing drastically 
the previous expressions~(\ref{covD3}).\footnote{In fact,
equations~\eqref{newcov} suggest to think of
${\cal Z}_{\mu}$ as some {\em composite connection} within the full affine algebra.}
Now, the fact that ${\cal X}_{\mu\nu}=0$ is a direct consequence of~\eqref{newcov} and
\bea
[\widehat{\cal D}_{\mu},\widehat{\cal D}_{\nu}]\,\hV &=&
\widehat{\cal H}^{{\cal A}}_{\mu\nu}\;T_{{\cal A}}\cdot\hV \;,
\eea
where $\widehat{\cal H}_{\mu\nu}$ is the field strength associated with the full connection
(\ref{fullcov}).

Summarizing, we have shown that under gauge 
transformations~\eqref{gaugescalar}, \eqref{gaugevector}
the Lagrangian ${\cal L}_{0}={\cal L}_{{\rm kin}}+{\cal L}_{{\rm top}}$ 
remains invariant up to total derivatives.
The local gauge algebra is spanned by 
generators $X_{{\cal M}}$~\eqref{genX} and is a subalgebra of the 
global symmetry algebra $\mathfrak{G}$ of the ungauged theory.
In particular, the gauge algebra may include hidden symmetries which in the 
ungauged theory are realized only on-shell.

\section{Gauge fixing}

In the previous section we constructed the deformation of the ungauged 
Lagrangian \eqref{EffLconfD2} 
that is invariant under the local version of a subalgebra 
of the affine symmetry algebra~$\mathfrak{G}$ of~\eqref{EffLconfD2}. 
The gauged Lagrangian has been obtained
by coupling vector fields with minimal couplings in covariant derivatives~\eqref{Lkin} 
and adding a topological term~\eqref{Ltop}. The gauging is entirely parametrized in terms of the 
embedding tensor $\Theta_{{\cal M}}$ which in particular encodes the local gauge algebra
with generators~\eqref{genX}.

With the new gauge fields and
a number of dual scalar fields the gauged Lagrangian contains more
fields than the original one, however as the new fields couple topologically only
they do not introduce new degrees of freedom.
More specifically, these fields arise with the first order field 
equations~\eqref{eomgauge} below, 
such that the additional local symmetries precisely eliminate the additional 
degrees of freedom.
In this section, we illustrate the various ways of gauge fixing the action and 
discuss the resulting different equivalent formulations of the theory.
Before that, we describe the generic properties of the scalar potential
which completes the construction of the bosonic sector of gauged supergravity.

\subsection{Scalar potential and equations of motion}

An important additional feature of gauged supergravity theories
is the presence of a scalar potential $V$ which is enforced in order to 
maintain supersymmetry of the deformed Lagrangian.
Its explicit form depends on the particular ungauged theory,
in particular on the number of supercharges. 
It must thus be computed case by case 
in the various supersymmetric theories
and we leave this for future work. 
Here we will just summarize the generic properties of this potential
and discuss their consequences for the gauged theory.
As a general property, the potential arises quadratic in the coupling constant $g$,
i.e.\ the deformed Lagrangian is supplemented by a term ${\cal L}_{{\rm pot}}=-g^{2}\,V$
where $V$ is bilinear in $\Theta_{{\cal M}}$, and generically depends on all scalar fields 
$\rho$, $\tilde\rho$, ${\cal V}$, $\hV$, and $\sigma$.
This dependence is constrained in order that its variation
takes the specific form
\bea
\delta V &=&
\frac{\delta V}{\delta \rho}\,\delta\rho 
+ \frac{\delta V}{\delta \sigma}\,\delta\sigma 
+ {\rm tr}\,\Big(\frac{\delta V}{\delta \Sigma}\,[{\cal V}^{-1}\,\delta {\cal V}]_{\mathfrak{p}}\Big)
+   \frac{\delta V}{\delta \hat\Sigma^{{\cal A}}}\,\delta \hat\Sigma^{{\cal A}}
\;,
\label{deltaV}
\eea
with $\frac{\delta V}{\delta \Sigma}\in\mathfrak{p}$, 
$\delta \hat\Sigma^{{\cal A}}\in\mathfrak{G}$ 
from~\eqref{varAS}.
Furthermore the various variations of $V$
are constrained such that~\eqref{deltaV} 
vanishes for gauge transformations~\eqref{gaugescalar},
i.e.\ the scalar potential is separately gauge invariant.
In particular, no further constraints on the 
embedding tensor will arise from its presence.

The total Lagrangian of the gauged theory then reads
\bea
{\cal L}&=& {\cal L}_{{\rm kin}}+{\cal L}_{{\rm pot}}+{\cal L}_{{\rm top}}
\label{Ltotal}
\\[2ex]
&=&
   \partial^\mua\! \rho \, {\cal D}_\mua \sigma - \ft 1 2 \, \rho \, \tr ( {\cal P}_\mua {\cal P}^\mua ) 
   -g^{2}\,V
\nonumber\\[1ex]
&&{}  -g\,\epsilon^{\mua\mub} \;\Big\{ 
    \tr\, \Big\langle \hat{\cal A}_{\mua}\, 
  \Big(\hV^{-1}(\partial_{\mub}\hV-\partial_{w}\hV\, \partial_\mub \tilde \rho )
      - Q_{\nu}
      - \frac{1+\gamma^{2}}{1-\gamma^{2}}\,P_{\mub}
      \Big) \Big\rangle_{\!w}\,
      -  {\cal A}^{(0)}_{\mua}\,\partial_\mub \tilde \rho\; \Big\}
     \nonumber \\ &&
                - \ft 1 2 \, g^2 \,\epsilon^{\mua\mub}\, {\cal A}^{(0)}_\mua {\cal A}^{(1)}_\mub   
                - \ft 1 2 \, g^2 \,\epsilon^{\mua\mub}\,
\tr\, \Big\langle \!\Big \langle \;
\frac 1 {v-w}\, [\hat {\cal A}_\mua(w)]_{\mathfrak k} \, [\hat {\cal A}_\mub(v)]_{\mathfrak k}
\, \Big \rangle_{\!v} \:\Big \rangle_{\!w}  \nonumber \\[1ex] &&
- \ft 1 2 \, g^2 \,\epsilon^{\mua\mub}\,
\tr\, \Big\langle \!\Big \langle \;
 \frac {(\gamma(v)-\gamma(w))^2 + (1-\gamma(v)\gamma(w))^2} {(v-w) (1-\gamma^{2}(v)) (1-\gamma^{2}(w))}\:
[\hat {\cal A}_\mua(w)]_{\mathfrak p} \, [\hat {\cal A}_\mub(v)]_{\mathfrak p}
                      \,  \Big \rangle_{\!v} \:\Big \rangle_{\!w}   \;.
                        \nonumber
\eea
\smallskip

\noindent
It gives rise to the following equations of motion:
\bea
&&
\partial_\mua \partial^\mua \rho = -g^{2}\,\frac{\delta V}{\delta \sigma}
\;,\qquad
\widehat{\cal D}_\mua {\cal D}^\mua \sigma =- \ft 1 2 \tr {\cal P}_\mua {\cal P}^\mua  
  -g^{2}\,\frac{\delta V}{\delta \rho}
\;,\qquad
\widehat{\cal D}_\mua (\rho {\cal P}^\mua) = g^{2}\, \frac{\delta V}{\delta \Sigma}\;,
\nonumber\\[1em]
&&
T_{\cAa,\cMa}{}^\cMb \,  \Theta_\cMb \, {\cal Z}^\cAa_\mua = 0 \;,\qquad
T_{\cAa,\cMa}{}^\cMb \, \widehat{\cal F}_{\mua\mub}^\cMa \, \Theta_\cMb =
   -2g\, \frac{\delta V}{\delta \hat\Sigma^{{\cal A}}}\;.
   \label{eomgauge}
   \eea
The duality equation $T_{\cAa,\cMa}{}^\cMb \,  \Theta_\cMb \, {\cal Z}^\cAa_\mua=0$
is not affected by the presence of the scalar potential while all other equations change.
In particular, a vanishing field strength is in general no longer compatible with the field equations,
i.e.\ the gauge fields have a nontrivial effect despite the fact that they are non-propagating
in two dimensions.
Note further, that the full covariant derivatives $\widehat{\cal D}_{\mu}$ defined in~\eqref{fullcov}
contain nontrivial ${\cal Z}^\cAa_{\mu}$ contributions even on-shell, 
as only the $\Theta$-projection of ${\cal Z}^\cAa_{\mu}$ vanishes by the equations of motion.

\subsection{Gauge fixing}

As anticipated above, the new fields $\hV$, $A_{\mu}^{\cal M}$ entering the gauged Lagrangian
induce first order equations of motion~\eqref{eomgauge}. 
Together with the additional local symmetry this implies 
that no new degrees of freedom are present in the gauged Lagrangian. 
In order to make this manifest, it may be useful to gauge-fix the local symmetry.
Also in order to make contact with the theories arising from particular compactification scenarios,
it will often be required to fix part of the extra local gauge symmetry, thereby effectively reducing the 
number of fields.
In this subsection we will discuss various ways of gauge fixing the action~\eqref{Ltotal}.

Let us first illustrate the relevant structures with an extremely simple toy example,
we consider the Lagrangian
\bea
{\cal L}&=& -\ft12\,\partial_{\mu}\varphi\,\partial^{{\mu}}\varphi
\;,
\label{Lfree}
\eea
of a free scalar field. The global shift symmetry $\varphi\rightarrow\varphi+c$
can be gauged by introducing covariant derivatives 
${\cal D}_{\mu}\varphi\equiv\partial_{\mu}\varphi-gA_{\mu}$.
The analogue of the full gauged Lagrangian~\eqref{Ltotal} then carries
a gauge field $A_{\mu}$ as well as a dual scalar field $\chi$ and
is of the form
\bea
{\cal L}&=& -\ft12\,{\cal D}_{\mu}\varphi\,{\cal D}^{{\mu}}\varphi
-g^{2}\,V(\chi)
-g\epsilon^{\mu\nu}A_{\mu}\,\partial_{\nu}\chi 
\;,
\label{Ltoy}
\eea
with the three terms representing the kinetic, the potential, and the topological term, respectively.
This action is obviously invariant under 
$\delta\varphi=g\lambda(x)\,,\;\delta A_{\mu}=\partial_{\mu}\lambda(x)$,
in particular, this restricts the potential $V$ to depend on the dual scalar field $\chi$ only.
The equation of motion derived from~\eqref{Ltoy} are
\bea
\partial^{\mu}{\cal D}_{\mu}\varphi =0\;,\qquad
{\cal D}_{\mu}\varphi=\epsilon_{\mu\nu}\,\partial^{\nu}\chi\;,\qquad
F_{\mu\nu}=g\epsilon_{\mu\nu}\,V'(\chi)\;,
\label{simple}
\eea
where the first equation consistently coincides with the integrability condition 
of the second equation.
There are (at least) three different ways of fixing the gauge freedom 
in~\eqref{Ltoy}.
\begin{itemize}

\item[i)] In the case of a vanishing potential $V=0$, and on a topologically trivial background, 
the vector field is pure gauge and may be put to zero, yielding the original Lagrangian~\eqref{Lfree}.
In this case, the deformation \eqref{Ltoy} thus is just a reformulation of the original model.

\item[ii)] For arbitrary potential $V$, the duality equation can be used to express $A_{\mu}$ in 
terms of scalar currents. On the Lagrangian level this leads to a theory expressed exclusively in terms of the dual
scalar field $\chi$
\bea
{\cal L}_{(1)}&=& -\ft12\,\partial_{\mu}\chi\,\partial^{{\mu}}\chi
-g^{2}\,V(\chi)
\;.
\eea
According to the reasoning of i), in the absence of a scalar potential 
this provides a dual formulation of the original model~\eqref{Lfree}.
This is (trivial) T-duality for the free scalar field. For more complicated systems the very same procedure yields
the known T-duality rules in the Abelian and the non-Abelian case~\cite{Buscher:1987sk}.
For non-vanishing potential, we obtain an equivalent formulation of the 'gauged' theory \eqref{Ltoy}
in which the kinetic term is replaced by a T-dual version in terms of dual scalar fields, in which no gauge fields are present.
The theory is in general no longer equivalent to the the original Lagrangian~\eqref{Lfree}
due to the presence of the scalar potential in order $g^{2}$.

\item[iii)]
For a quadratic potential $V(\chi)=V_{0}+\frac12m^{2}\chi^{2}$, i.e.\ considering the lowest order 
expansion around a stationary point, the equations of motion may be used to
replace $mg\chi=F_{\mu\nu}$. Simultaneously fixing the gauge freedom by setting $\varphi=0$,
one arrives at a Lagrangian
\bea
m^{2}{\cal L}_{(2)} &=& -\ft14F^{\mu\nu}F_{\mu\nu} - \ft12g^{2}m^{2}\,A_{\mu}A^{\mu}-g^{2}m^{2}V_{0}\;,
\eea
of a massive vector field which now carries the degree of freedom of the system. 
This is the standard Higgs mechanism in two dimensions.

\end{itemize}

Gauge fixing of the general Lagrangian~\eqref{Ltotal} is considerably more complicated due to the 
high nonlinearity of the system, but schematically follows precisely the same pattern.
In applications to describe the effective actions of concrete compactifications with
non-vanishing cosmological constant, the last procedure iii) will be often the most appropriate
one in order to identify the correct distribution of the degrees of freedom among different 
supermultiplets.
{}From a  point of view, the gauge fixing according to ii) is the most interesting.
In the context of the full model~\eqref{Ltotal} it extends to the following:
the duality equations $T_{\cAa,\cMa}{}^\cMb \,  \Theta_\cMb \, {\cal Z}^\cAa_\mua = 0$
can be solved as algebraic equations for the vector fields 
$\Theta_{{\cal M}}{}^{{\cal A}}A_{\mu}^{{\cal M}}$.
The explicit formulas may be arbitrarily complicated of course.
Plugging this back into the Lagrangian leads to an equivalent formulation
of the model in which the vector fields have been completely removed from the action.
As in ii) this exchanges the kinetic term by a T-dual version in terms
of dual scalar fields.
In this formulation the only effect of the gauging is the scalar potential 
which remains unaffected by the gauge fixing.
We conclude that for every gauging in two dimensions there is a formulation in 
a T-dual frame, i.e.\ a formulation in terms of a combination of original and dual scalars, 
in which no gauge fields enter the Lagrangian and the only effect of the gauging is the scalar potential.
(In general, this will not be the most convenient frame to identify a particular
higher-dimensional origin.)

Let us consider as an example a gauging in which a subalgebra of the
zero-modes of $\hat{\mathfrak{g}}$, i.e.\ of the algebra of target-space isometries $\mathfrak{g}$ is gauged.
According to~\eqref{Ltop} this will induce a topological term which couples the
gauge fields to the (algebra-valued) dual potentials $Y_{1}$.
No higher dual potentials enter the Lagrangian.
Apart from some additional subtleties related to the coset structure of~\eqref{Lkin},
the resulting couplings are precisely of the type 
considered in \cite{Hull:1989jk}. Integrating out the vector fields
in absence of a scalar potential gives rise to a dual formulation of the model
and reproduces the known formulas of non-Abelian 
T-duality~\cite{Buscher:1987sk,Hull:1989jk,delaOssa:1992vc,Giveon:1993ai,
Alvarez:1993qi,Alvarez:1994zr,Mohammedi:1995mf}.
In particular, since (in contrast to the simplified example~\eqref{simple}) 
the duality equations in this case carry the vector fields 
on both sides, the procedure gives rise to antisymmetric couplings
$\epsilon^{\mu\nu}\,\partial_{\mu}Y_{1}{}^{\!\alpha}\,\,
\partial_{\nu}Y_{1}{}^{\!\beta}\,\,B_{[\alpha\beta]}$
among the dual scalar fields in the new frame.
For maximal supergravity, an example of different scalar frames has
been worked out in~\cite{Fre':2005si}.

As discussed above, the gauge groups appearing in 
our construction~\eqref{Ltotal} will in general go beyond 
the off-shell symmetry of the ungauged theory, i.e.\ beyond the
target-space isomorphisms of the original $\sigma$-model.
They will thus naturally lead to a far broader class of equivalent formulations
of the kinetic sector, obtained after integrating out the vector fields. 
The proper framework to systematically incorporate these different formulations
is presumably Lie-Poisson T-duality,
see~\cite{Klimcik:1995ux,Klimcik:1995jn,Sfetsos:1997pi,Stern:1999dh}.
We defer a systematic treatment to future work.
Let us stress once more that due to the presence of a scalar potential,
the gaugings~\eqref{Ltotal} describe genuinely
inequivalent deformations of the ungauged Lagrangian~\eqref{EffLconfD2}.

\section{Maximal supergravity}
\label{Sec:maximal}

One of the richest examples in two dimensions is the 
theory obtained by dimensional reduction from eleven-dimensional
supergravity giving rise to maximal $N=16$ supergravity with scalar coset space 
${\rm G}/{\rm K}={\rm E}_{8(8)}/{\rm SO}(16)$ as a particular case of the integrable structures introduced 
above~\cite{Juli85,Nicolai:1987kz,Nicolai:1988jb,Nicolai:1998gi}.
The symmetry of the ungauged theory is the affine algebra
$\mathfrak{e}_{9(9)}\equiv\widehat{\mathfrak{e}_{8(8)}}$.
In this section we will illustrate with a number of examples 
the general construction of gaugings in two dimensions
starting from the maximal theory.
In subsection~\ref{Sec:E8} we describe gaugings that are naturally
formulated in the $\mathfrak{e}_{8}$ grading of~$\mathfrak{e}_{9(9)}$. These have
a natural interpretation as reductions from three-dimensional supergravity theories.
In subsection~\ref{Sec:SL9} we describe gaugings in the $\mathfrak{sl}(9)$
grading of~$\mathfrak{e}_{9}$, these include the ${\rm SO}(9)$ gauging
corresponding to an $S^{8}$ compactification of the ten-dimensional IIA theory
as well as flux gaugings from eleven dimensions.
Gaugings with type IIB origin are discussed in subsection~\ref{other}.

\subsection{The basic representation of ${\rm E}_{9}$}

In order to construct the gaugings of the maximal ${\rm E}_{8(8)}/{\rm SO}(16)$ 
theory
the first task is the choice of representation of vector fields used in the gauging.
Extrapolating the representation structures from higher dimensions it turns out 
that the relevant representation for the gauge fields
is the basic representation of $\mathfrak{e}_{9(9)}$,
i.e.\ the unique level~1 representation of this affine algebra.
In the following we will see more specifically that the basic representation
reproduces precisely the structures expected from dimensional reduction;   
the complete proof will ultimately include consistency with the supersymmetric extension.

Branching the basic representation of $\mathfrak{e}_{9(9)}$ under $\mathfrak{e}_{8}$, the vector fields hence
transform as
\bea
{\rm basic}&\rightarrow&
{\bf 1}_{0}\,\oplus\,\nonumber\\&&{}
{\bf 248}_{-1}\,\oplus\, \,\nonumber\\&&{}
({\bf 1}\!\oplus\!{\bf 248}\!\oplus\!{\bf 3875})_{-2}\,\oplus\,
\nonumber\\&&{}
({\bf 1}\!\oplus 2\!\cdot\!{\bf 248}\oplus\!{\bf 3875}\!\oplus\!{\bf 30380})_{-3}\,\oplus\,
\nonumber\\&&{}
(2\!\cdot\!{\bf 1}\oplus3\!\cdot\!{\bf 248}\oplus2\!\cdot\!{\bf 3875}\oplus\!{\bf 30380}
\!\oplus\!{\bf 27000}\!\oplus\!{\bf 147250})_{-4}
\,\oplus\,\dots
\;,
\label{charE8}
\eea
where the subscript denotes the $L_{0}$ charge of the associated Virasoro algebra.
The embedding tensor $\Theta_{{\cal M}}$ transforms in the conjugate vector field representation, 
i.e.\ its components carry $L_{0}$ charges opposite to~(\ref{charE8}).
Counting the $L_{0}$ charge in powers of a variable $y$,
the character of the basic representation of $\mathfrak{e}_{9}$
is given by the famous McKay-Thompson series
\bea
\chi_{\omega_{0}}(y) = j^{1/3}(y)=
1+248\,y+4124\,y^{2}+34752\,y^{3}+213126\,y^{4}+1057504\,y^{5}+\dots
\;,
\label{char}
\eea
in terms of the modular invariant $j(y)$~\cite{MR563927,MR604635}.
The symmetric product~(\ref{prodvec}) takes the form~\cite{DiFrancesco:1997nk}
\bea
\chi_{\omega_{0}}(y) \otimes_{{\rm sym}} \chi_{\omega_{0}}(y)
&=&
\chi^{{\rm vir}}_{(1,1)}(y)\,\chi_{2\omega_{0}}(y)+
\chi^{{\rm vir}}_{(2,1)}(y)\,\chi_{\omega_{7}}(y)
\;,
\label{char2}
\eea
where $\chi_{2\omega_{0}}$ and $\chi_{\omega_{7}}$ denote the
characters of the level 2 representations starting from 
a ${\bf 1}$ and a ${\bf 3875}$ of $\mathfrak{e}_{8}$, respectively.
As discussed in section~\ref{sec:qc} above, 
the multiplicities $\chi^{{\rm vir}}_{(1,1)}$, $\chi^{{\rm vir}}_{(2,1)}$
carry representations of the coset CFT with central charge given by~(\ref{CFTcentral}),
which in this case yields $c=1/2$, i.e.\ the Ising model. Accordingly
\bea
\chi^{{\rm vir}}_{(1,1)}(y) &=&
1+y^2+y^3+2y^4+2y^5+\dots \;,\nonumber\\
\chi^{{\rm vir}}_{(2,1)}(y) &=&
1+y+y^2+y^3+2y^4+2y^5+\dots \;,
\label{charVir}
\eea
denote the lowest $c=1/2$ Virasoro representations.
Consistent gaugings of two-dimensional maximal supergravity
thus correspond to components within the expansion~(\ref{char})
such that their two-fold symmetric product is sitting in a quasi-primary state
of~(\ref{charVir}) on the r.h.s.\ of~(\ref{char2}).
In principle, all gaugings can be determined this way. In the next subsections
we work out a few examples.

\subsection{Gaugings in the ${\rm E}_{8}$ grading}
\label{Sec:E8}

\begin{figure}[tbp]
   \centering
   \includegraphics[bb=0 0 600 800,width=8cm,angle=270]{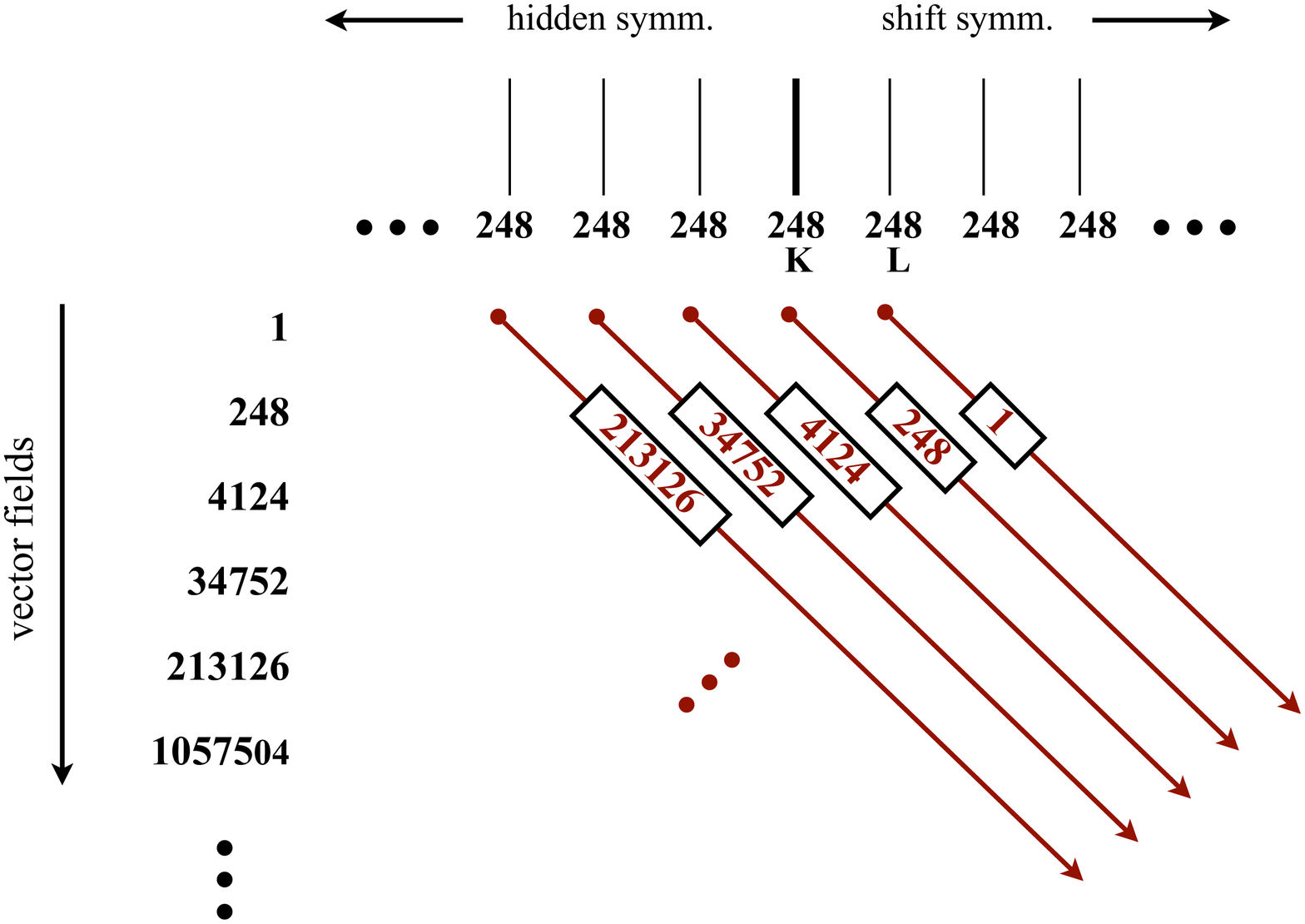}
   \caption{Couplings induced by different components of the embedding tensor $\Theta_{{\cal M}}$.}
   \label{Figure:theta}
\end{figure}

According to (\ref{ConstraintLinear}), the embedding tensor $\Theta$ transforms 
in the conjugate vector field representation. It describes the couplings of vector fields
to $\mathfrak{e}_{9(9)}$ symmetry generators according to~(\ref{covD})
\bea
{\cal D}_\mua &=
\partial_\mua - g \, {\cal A}_\mua^{{\cal M}} \, 
\Theta_{{\cal M}}{}^{{\cal A}}\,T_{{\cal A}} \;.
\label{covDA}
\eea
It is instructive to visualize these couplings as in Figure~\ref{Figure:theta}.
The $\mathfrak{e}_{9(9)}$ symmetry generators are plotted horizontally 
with the $L_{0}$ charge increasing from left to right, the vector fields are plotted vertically.
The diagonal lines represent the couplings induced by each component of $\Theta$.
The figure shows that every gauging defined by a particular component of $\Theta$ involves only a finite 
number of hidden and zero-mode symmetries 
and an infinite tower of unphysical shift symmetries. 
As discussed above this implies in particular that only the finite number of vector fields 
coupled to the physical symmetries appears in the Lagrangian.

The simplest gauging in this description is defined by the lowest $\Theta$ component 
in the basic representation, i.e.\ by the highest weight singlet ${\bf 1}_{0}$ in~(\ref{charE8}).
According to Figure~\ref{Figure:theta} this is a gauging of only shift symmetries.
As a consequence, the quadratic constraint is automatically satisfied
as can be seen from its form~(\ref{Quadratic2}), such that this component indeed represents a consistent gauging.
Moreover, as only unphysical symmetries are involved, the gauging 
will be invisible in the kinetic and topological part ${\cal L}_{\rm kin}+{\cal L}_{\rm top}$
of the Lagrangian.
Its only contribution to the total Lagrangian~\eqref{Ltotal} is via the scalar potential $V$.
This gauging has in fact a simple higher-dimensional origin descending from
dimensional reduction of the three-dimensional maximal ungauged theory~\cite{Marcus:1983hb}.
With the ansatz
\be
e_m{}^a ~=~ \left(\begin{array}{cc} 
\delta_\mu^\alpha\,e^{\lambda} & \rho B_\mu \\
0&\rho \end{array} \right)
\;,\qquad
m,a\in\{1,2,3\}\;,
\quad
\mu,\alpha\in\{1,2\}\;,
\ee
for the three-dimensional vielbein in terms of 
a conformal factor $\lambda$, dilaton $\rho$ and Kaluza-Klein
vector field $B_{\mu}$, the three-dimensional Einstein field equations give rise to
\bea
\partial^{\mu}(\rho^{3}\lambda^{-2}\partial_{[\mu}B_{\nu]})&=& 0
\;,
\eea
which is solved by $\partial_{[\mu}B_{\nu]}=\rho^{-3}\lambda^{2}C\,\epsilon_{\mu\nu}$
with a constant $C$. The ungauged two-dimensional theory is obtained by setting $C=0$. In contrast,
keeping a non-vanishing $C$ and thus a non-vanishing field-strength of the Kaluza-Klein
vector field precisely corresponds to the singlet gauging induced by the lowest components of $\Theta$.
In accordance with the above observations
the only effect of $C$ in the Lagrangian is the creation of a 
scalar potential $\rho^{-3}\lambda^{3}C^{2}$
descending from the kinetic term ${\cal L}_{B}\propto
\partial_{[\mu}B_{\nu]}\partial^{\mu}B^{\nu}$.
As discussed after equation~(\ref{L1action}) the effect of
this scalar potential is a deformation of the free field equation 
satisfied by the dilaton $\rho$ which necessitates gauging of the $L_{1}$ shift symmetry by the Kaluza-Klein
vector field $B_{\mu}$.
This is precisely the lowest coupling exhibited in Figure~\ref{Figure:theta}.

At the next level in $\Theta$ comes the ${\bf 248}_{1}$. 
According to Figure~\ref{Figure:theta}, the corresponding gaugings involve
apart from the infinite tower of unphysical symmetries
a single generator of the $\mathfrak{e}_{8}$ zero-modes which couples to the 
Kaluza-Klein vector field. Again one verifies that the quadratic constraint is automatically satisfied. 
These are precisely the Scherk-Schwarz gaugings~\cite{Scherk:1979zr,Andrianopoli:2002mf,deWit:2002vt} 
obtained from three dimensions,
singling out one among the generators of the
global symmetry algebra  $\mathfrak{e}_{8}$ in three dimensions.

At the third level, $\Theta$ has three components ${\bf 1}_{2}$, ${\bf 248}_{2}$, ${\bf 3875}_{2}$.
As can be seen from Figure~\ref{Figure:theta}, the gaugings induced by the ${\bf 248}_{2}$
for the first time involve the hidden symmetries $T_{\alpha,-1}$ coupled to the Kaluza-Klein vector field. 
Those gaugings described by the
${\bf 1}_{2}\oplus{\bf 3875}_{2}$ on the other hand 
involve only the $\mathfrak{e}_{8}$ zero-mode symmetries
coupled to the  ${\bf 248}_{-1}$ vector fields.
These are the theories obtained by dimensional reduction of the 
three-dimensional maximal gauged theories
described by an embedding tensor in precisely this representation~\cite{Nicolai:2000sc}.
For all these theories there is a nontrivial quadratic constraint to be satisfied by 
the components of~$\Theta$.

To summarize, all the gaugings with three-dimensional origin are naturally identified
within~Figure~\ref{Figure:theta}. The lowest components of the vector fields
in the expansion~(\ref{charE8}) correspond
to the Kaluza-Klein vector field ${\bf 1}_{0}$ and the vector fields ${\bf 248}_{-1}$
descending from the three-dimensional vector fields, respectively. 
Higher components of the embedding tensor involve higher hidden 
symmetries and increasingly nontrivial quadratic constraints. 
A priori, it is not clear if there are nontrivial solutions of
the quadratic constraint that involve
arbitrarily high components of $\Theta$ in the 
expansion~(\ref{charE8}).
The higher-dimensional origin of the associated gaugings remains to be elucidated.

\subsection{Gaugings in the ${\rm SL}(9)$ grading}
\label{Sec:SL9}

By far not all gaugings of two-dimensional maximal supergravity 
have a natural place in~Figure~\ref{Figure:theta}. 
Although all of them 
can be identified among the components of the expansion~(\ref{char})
of the embedding tensor $\Theta_{{\cal M}}$,
the major part will be hidden at higher levels and in linear combinations
of these components.
In some cases it may however be possible to naturally identify them 
within other gradings of the affine algebra.
As an example we will present in this section the 
theory obtained by dimensional reduction of the IIA theory
on a (warped) eight-sphere $S^{8}$~\cite{Boonstra:1998mp,Nicolai:2000zt,Bergshoeff:2004nq},
which plays a distinguished role in (a low dimensional version of)
the AdS/CFT correspondence~\cite{Itzhaki:1998dd,Boonstra:1998mp,Youm:1999dc}.
Its gauge group contains an $SO(9)$ as its semisimple part.
Closely related are the compactifications on the non-compact manifolds $H^{p,8-p}$ that result in gauge
groups ${\rm SO}(p,9-p)$. We will identify the embedding tensors
$\Theta_\cMa$ that define these theories.

These gaugings are most conveniently described in the $\mathfrak{sl}(9)$
grading of $\mathfrak{e}_{9(9)}$. The intersection of zero-modes of this grading
and the $\mathfrak{e}_{8}$ grading of the previous section is given by
\bea
\mathfrak{e}_{8(8)} \,\cap\,\mathfrak{sl}(9) &=& \mathfrak{sl}(8) \oplus \mathfrak{gl}(1)
\;.
\label{intersect}
\eea
Denoting by $\ell_{\mathfrak{e}8}$ and $\ell_{\mathfrak{sl}\,9}$ the charges associated with the 
$\mathfrak{e}_{8}$ and the $\mathfrak{sl}(9)$ grading, respectively,
they are related by
\bea
\ell_{\mathfrak{sl}\,9}&=& \ell_{\mathfrak{e}8} + q \;,
\eea
where $q\in \frac13{\mathbb Z}$ is the charge associated with the $\mathfrak{gl}(1)$
factor in~(\ref{intersect}). E.g.\ the level $\ell$ in the $\mathfrak{e}_{8}$ grading of the adjoint representation
decomposes as
\bea
   {\bf 248}_\ell   
& \rightarrow &  {\bf 8}'_{\ell+1} \,\oplus\, {\bf 28}_{\ell+2/3} \,\oplus\, 
{\bf 56}'_{\ell+1/3}
                             \,\oplus\, {\bf 1}_{\ell} \,\oplus\, {\bf 63}_{\ell} \,\oplus\, {\bf 56}_{\ell-1/3} 
                             \,\oplus\, 
{\bf 28}'_{\ell-2/3} \,\oplus\, {\bf 8}_{\ell-1}
\;,
\nonumber
\eea
under $\mathfrak{sl}(8)$ where the subscript on the r.h.s.\ indicates $\ell_{\mathfrak{sl}\,9}$.
This shows in particular that the $\mathfrak{sl}(9)$ algebra building the zero-modes in this grading
is composed out of the ${\bf 8}'$, ${\bf 1}\oplus\, {\bf 63}$, and ${\bf 8}$
with $\ell_{\mathfrak{e}8}$ charges $-1$, $0$, and $1$, respectively.
The adjoint representation in the $\mathfrak{sl}(9)$ grading takes the well known form
\bea
{\rm adj}&\rightarrow&
\dots\,\oplus\,{\bf 80}_{-1}\,\oplus\, {\bf 84}'_{-2/3}\,\oplus\, {\bf 84}_{-1/3}\,\oplus\,
{\bf 80}_{0}\,\oplus\,{\bf 84}'_{1/3}\,\oplus\, {\bf 84}_{2/3}\,\oplus\,{\bf 80}_{1}\,\oplus\,
\dots\;.
\nonumber\\
\label{adjSL9}
\eea
Similarly, one computes the form of the basic representation~(\ref{charE8})
in the $\mathfrak{sl}(9)$ grading which gives rise to
\bea
{\rm basic}&\rightarrow&
{\bf 9}'_{0}\,\oplus\,\nonumber\\&&{}
{\bf 36}_{-1/3}\,\oplus\,\nonumber\\&&{}
{\bf 126}'_{-2/3}\,\oplus\,\nonumber\\&&{}
({\bf 9}'\oplus{\bf 315})_{-1}\,\oplus\,\nonumber\\&&{}
({\bf 36}\oplus{\bf 45}\oplus{\bf 720}')_{-4/3}
\,\oplus\,\dots
\;.
\label{charSL9}
\eea
It is instructive to note that the parts with coinciding 
$(\ell_{\mathfrak{sl}\,9}\;{\rm mod}\;1)$ in~(\ref{charSL9})
constitute the three irreducible representations under the
$\widehat{\mathfrak{sl}(9)}$ subalgebra of~(\ref{adjSL9})
(this can be inferred, for example,
from the decompositions given in \cite{Kac:1988iu}).

With the vector fields decomposed as~(\ref{charSL9}),
it is straightforward to identify the eleven-dimensional origin of the lowest components.
These are the Kaluza-Klein vector (${\bf 9}'_{0}$),
the vector fields that originate from the three-form (${\bf 36}_{-1}$) and 
the vector fields coming from the dual six-form (${\bf 126}'_{-2}$) of eleven-dimensional supergravity. 
A priori, a possible eleven-dimensional origin of the higher components remains unclear.
Note however, that we have already
identified a higher-dimensional origin for different vector fields than in the reduction from
three dimensions discussed in the previous section.
Analysis of more complicated dimensional reductions may disclose a higher-dimensional
origin of yet other vector fields within the basic representation of $\mathfrak{e}_{9(9)}$. 

\begin{figure}[tbp]
   \centering
   \includegraphics[bb=0 0 600 800,width=8cm,angle=270]{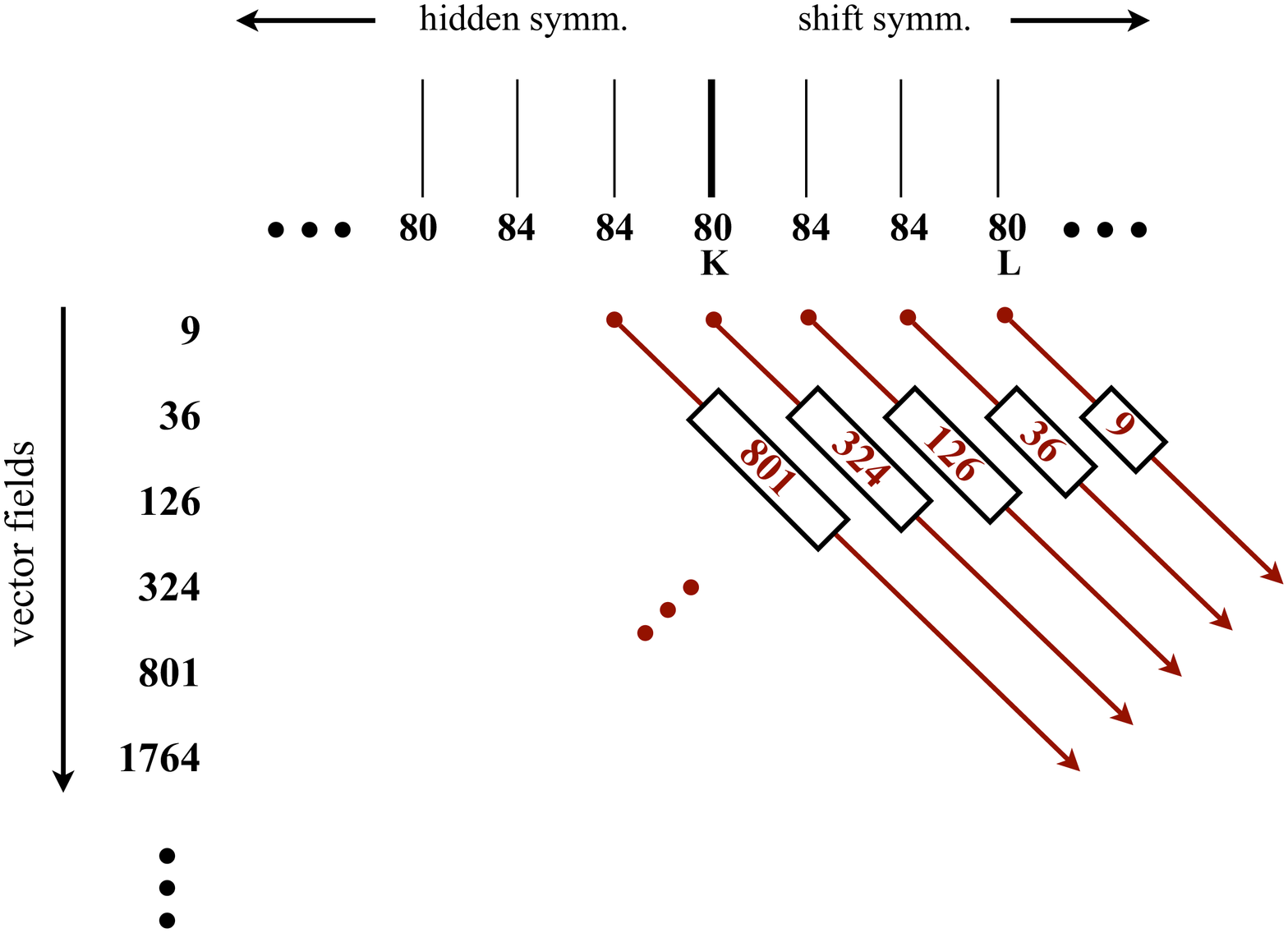}
   \caption{Couplings induced by different components of the embedding tensor $\Theta_{{\cal M}}$.}
   \label{Figure:theta9}
\end{figure}

The embedding tensor $\Theta_\cMa$ transforms in the conjugate vector field representation.
Accordingly, we may try to identify the gaugings
associated with the various components of $\Theta$ in the expansion conjugate to~(\ref{charSL9}).
The induced couplings are schematically depicted in Figure~\ref{Figure:theta9}.
Similar to the discussion in the previous section, the lowest components 
${\bf 9}_{0}$,  ${\bf 36}'_{1/3}$, ${\bf 126}_{2/3}$ correspond to nontrivial fluxes
associated with the vector fields in the reduction from eleven dimensions.
As manifest in the figure, these gaugings involve only shift symmetries in the
$\mathfrak{sl}(9)$ grading.

We will be interested by the gaugings induced by the ${\bf 45}'_{4/3}$.
With a little effort one may show that an embedding tensor in this representation
automatically satisfies the quadratic constraint~(\ref{Quadratic2}).
Namely, working out the couplings induced by this ${\bf 45}'_{4/3}$ in Figure~\ref{Figure:theta9},
it follows from the $\mathfrak{sl}(9)$  representation structure that the lowest symmetry generators
which are involved in the gauging are sitting in the ${\bf 80}_{0}$, the ${\bf 84}_{2/3}$, and the 
${\bf 80}_{1}$. 
In particular, the latter couple only to the ${\bf 45}_{-4/3}$ of the 
vector fields.\footnote{This can be seen as follows.
According to \eqref{covD} and \eqref{ConstraintLinear} the vector fields couple to generators as
${\cal A}_\mua^{{\cal M}} \,
(T_{\cAb,{\cal M}}{}^{{\cal N}}\,\eta^{\cAa\cAb}\, \Theta_{{\cal N}})
\,T_{{\cAa}}$. Since $\eta^{\cAa\cAb}$ is invariant under $L_1$,
indices in the range ${\cAa}\in{\bf 80}_1$ couple to ${\cAb}\in{\bf 80}_0$, i.e.\
in this case $T_\cAb$ is just the ${\rm SL}(9)$. Since
\eqref{charSL9} is a decomposition into irreducible ${\rm SL}(9)$
components and the indices '$_{\cal N}$' are in the range 
${\cal N}\in{\bf 45}'_{4/3}$ 
(as this is the only non-vanishing $\Theta$-component) 
the range of indices '$^{\cal M}$' is restricted to
${\cal M}\in{\bf 45}_{-4/3}$.
}
The form of the quadratic constraint~(\ref{Quadratic2}) then shows that its only nontrivial contribution
can sit in the component where {\footnotesize${\cal M}$} and {\footnotesize${\cal N}$} take
values in the ${\bf 36}'_{1/3}$ and the ${\bf 45}'_{4/3}$, respectively, i.e.\
live in the $\mathfrak{sl}(9)$ tensor product $ {\bf 36}' \otimes {\bf 45}' = 
{\bf 630}' \oplus {\bf 990}'$.
Since there is no overlap with the representations actually present in 
the square of this embedding tensor
(${\bf 45}' \otimes_{{\rm sym}} {\bf 45}' = 
{\bf 495}' \oplus {\bf 540}'$),
the quadratic constraint is automatically satisfied.
We have thus shown that an embedding tensor in the ${\bf 45}'_{4/3}$
defines a consistent gauging in two dimensions.
This representation can be parametrized by a symmetric $9 \times 9$ matrix $Y$. 
By fixing part of the ${\rm SL}(9)$ symmetry
this matrix can be brought into the form
\begin{align}
  Y &= {\rm diag}(\,\underbrace{1, \dots,}_{p}\underbrace{-1,\dots,}_{q} \underbrace{0, \dots}_{r}\,) \;,
\end{align}
with $p+q+r=9$. Such an embedding tensor gauges a subalgebra ${\mathfrak{cso}}(p,q,r)$ of the zero-mode 
algebra $\mathfrak{sl}(9)$ in ~\eqref{adjSL9}. The corresponding gauge fields come from the $ {\bf 36}_{-1/3}$. 
For $r=q=0$ this is the ${\rm SO}(9)$ gauging corresponding to the IIA $S^{8}$ compactification
mentioned above.
In addition there is the infinite tower of shift-symmetries accompanying this gauging,
starting from the full ${\bf 84}_{+2/3}$, a ${\bf 44}$ inside the ${\bf 80}_{+1}$, etc.

It is instructive to visualize this ${\rm SO}(9)$ gauging within the $\mathfrak{e}_{8}$ grading
of Figure~\ref{Figure:theta}. In that table, the ${\rm SO}(9)$ singlet component of $\Theta$
which defines the gauging is a linear combination of the two  ${\rm SO}(8)$ singlets appearing in the 
branching of the ${\bf 3875}_{2}$ and the ${\bf 147250}_{4}$ under ${\rm SO}(8)$.
In the $\mathfrak{e}_{8}$ grading this gauging thus involves a number of hidden and zero-mode symmetries.
More precisely, the gauge group appearing in the Lagrangian (\ref{Ltotal})
is of the non-semisimple form
\bea
   G = {\rm SO}(8) \ltimes 
   \Big((\mathbb{R}_+^{28} \times \mathbb{R}_+^{8} )_{0} \times (\mathbb{R}_+^8){}_{-1} \Big)
   \;,
\eea
with the $(\mathbb{R}_+^{28} \times \mathbb{R}_+^{8} )_{0}$, and $(\mathbb{R}_+^8)_{-1}$ 
corresponding to zero-mode symmetries and hidden symmetries from level $-1$, respectively.
{}From this perspective it is thus not at all obvious that an ${\rm SO}(9)$ gauge group is realized.
Instead, the ``off-shell gauge group'' involves the maximal Abelian ($36$-dimensional) subalgebra
of the zero-mode $\mathfrak{e}_{8}$.

\subsection{Other gradings}
\label{other}

The $SO(9)$ example presented in the last section 
already shows that particular gaugings may be far more transparent
within one grading than within another. It will thus be interesting to analyze the
gaugings manifest in the different gradings of 
$\mathfrak{e}_{9(9)}$. A table of the 112 
maximal rank subalgebras of $\mathfrak{e}_{8}$
corresponding to the zero-mode algebras in the different gradings 
can be found in~\cite{Hollowood:1987hf}.
Of particular interest may be the $\mathfrak{so}(8,8)$ grading giving rise
to a decomposition 
\bea
{\rm adj} &\rightarrow& \dots\,\oplus\,
({\bf 128_{s}})_{-1/2}\,\oplus\,
{\bf 120}_{0}\,\oplus\, ({\bf 128_{s}})_{1/2}\,\oplus\, {\bf 120}_{1}\,\oplus\,
\dots \;,\nonumber\\[2ex]
{\rm basic} &\rightarrow&
{\bf 16}_{0}\,\oplus\,\nonumber\\&&{}
({\bf 128_{c}})_{-1/2}\,\oplus\,\nonumber\\&&{}
({\bf 16}\oplus{\bf 560})_{-1}\,\oplus\,\nonumber\\&&{}
({\bf 128_{c}}+{\bf 1920_{s}})_{-3/2}\,\oplus\,
\dots
\;,
\label{adjSO88}
\eea
of the adjoint and the basic representation, respectively. This grading
is particularly adapted to identify the transformation behavior of the different $\Theta$ 
components (e.g.\ fluxes, twists, etc.) under the ${\rm SO}(8,8)$ duality group.

Another grading of interest is the one w.r.t.~$\mathfrak{sl}(8)\times \mathfrak{sl}(2)$
\bea
{\rm adj} &\rightarrow& 
\dots \,\oplus\,
{\bf (28',2)}_{-1/4}\,\oplus\,
({\bf (63,1)}\!\oplus\!{\bf (1,3)})_{0}\,\oplus\, 
{\bf (28,2)}_{1/4}\,\oplus\, {\bf (70,1)}_{1/2}\nonumber\\
&&{}
\,\oplus\,
{\bf (28',2)}_{3/4}
\,\oplus\,({\bf (63,1)}\!\oplus\!{\bf (1,3)})_{1}\,\oplus\,
\dots \;,\nonumber\\[2ex]
{\rm basic} &\rightarrow&
{\bf( 8', 1 )}_{0}\,\oplus\,\nonumber\\&&{}
{\bf( 8, 2 )}_{-1/4}\,\oplus\,\nonumber\\&&{}
{\bf( 56, 1 )}_{-1/2}\,\oplus\,\nonumber\\&&{}
{\bf( 56', 2 )}_{-3/4}\,\oplus\,\nonumber\\&&{}
({\bf( 8', 1\oplus3 )}\oplus(\bf{216,1}))_{-1}\,\oplus\,\nonumber\\&&{}
({\bf( 216', 2 )}\oplus2\!\cdot\!{\bf( 8, 2 )})_{-5/4}\,\oplus\,
\dots
\;,
\label{adjSL82}
\eea
related to the ten-dimensional IIB theory,
with $\mathfrak{sl}(8)$ and $\mathfrak{sl}(2)$ reflecting the torus $T^{8}$
and the IIB symmetry, respectively.
By regarding the representation content,
it is easy to verify that the lowest entries of the basic representation
in this grading correspond to the gaugings induced by IIB $p$-form and geometric fluxes on $T^{8}$.

%%%%%%%%%%%%%%%%%%%%%%%%%%%%%%%%%%%%%%%%%%%%%%%%%%%%%%%%%%%%
\section{Conclusions and outlook}
%%%%%%%%%%%%%%%%%%%%%%%%%%%%%%%%%%%%%%%%%%%%%%%%%%%%%%%%%%%%

In this paper, we have presented the construction of gaugings of two-dimensional supergravity.
We have shown how to consistently gauge subalgebras of the affine global symmetry algebra~$\mathfrak{G}$ 
of the ungauged theory by coupling vector fields in a highest weight representation of the affine algebra 
with a particular topological term~\eqref{Ltop}.
The gaugings are described group-theoretically in terms of a constant embedding tensor $\Theta_{{\cal M}}$
in the conjugate vector representation and subject to the quadratic consistency constraint~\eqref{Quadratic3}.
This tensor parametrizes the different theories, 
defines the gauge algebra and entirely encodes the gauged Lagrangian~\eqref{Ltotal}.
The resulting gauge algebras are generically infinite-dimensional and include hidden symmetries 
which are on-shell and not among the target-space isometries of the ungauged theory.
Yet, only a finite part of the gauge symmetry is realized on the Lagrangian level 
(with its infinite-dimensional tail exclusively acting on dual scalar fields that are not 
present in the Lagrangian) and only a finite number of gauge fields enters the Lagrangian.
As a main result, we have shown that the total Lagrangian~\eqref{Ltotal} is invariant under
the action~\eqref{gaugescalar}, \eqref{gaugevector} of the local gauge algebra.
In absence of a scalar potential, particular gauge fixing shows that the gauging, 
merely amounts to a (T-dual) reformulation of the ungauged theory.
A scalar potential on the other hand induces a genuine deformation of the original theory.
We have worked out a number of examples for maximal ($N=16$) supergravity
in two dimensions which illustrate the structure of the gaugings.
In particular, we have discussed the gaugings corresponding to those 
components of the embedding tensor with lowest charge with respect to several 
gradings of $\mathfrak{e}_{9(9)}$ which allow for a straightforward higher-dimensional interpretation. 
\smallskip

The presented construction opens up a number of highly interesting questions concerning its 
applications as well as possible generalizations.
E.g.~we have motivated the particular ansatz~\eqref{ConstraintLinear} for the embedding tensor
by the observation that it reduces the quadratic consistency constraints~(\ref{ConstraintQuadratic}) 
and (\ref{Quadratic2}) to the same equation (\ref{Quadratic3}).
Moreover, it seems in line with the findings in higher-dimensional theories that
the embedding tensor transforms in the dual representation of the ($D-1$)-forms in a given dimension~$D$.
Yet, it would be interesting to study, if the present construction could be generalized to more general 
choices of the embedding tensor. A related question is the particular choice of the vector field representation.
While the general bosonic construction seems to yield no preferred representation for the gauge fields 
(and thus for the embedding tensor) it is presumably consistency with the supersymmetric extension that
puts severe constraints on this choice.  

The analysis of this paper has been performed for a general two-dimensional bosonic
coset space sigma-model. Above all, it remains to extend the presented construction to the 
fermionic sector of the various supersymmetric theories.
Of particular interest is the maximal ($N=16$) supergravity theory. 
As the integrable structures of the ungauged bosonic theory naturally extend to the full 
theory~\cite{Nicolai:1987kz,Nicolai:1988jb,Nicolai:1998gi}
the construction should straightforwardly extend.
In particular, this should elucidate the role of the 
basic representation which we have found relevant for the maximal theory.
The construction will fix the fermionic mass terms and yield
the specific form of the scalar potential.
A crucial ingredient will be the representation structure of the
infinite-dimensional subalgebra ${\mathfrak k}(\mathfrak{e}_{9})$ of $\mathfrak{e}_{9(9)}$
under which the fermions transform~\cite{Nicolai:2004nv,Paulot:2006zp,Kleinschmidt:2006dy}.
What we have only started in section~5 of this paper is the study of the various resulting 
two-dimensional theories; this analysis needs to be addressed systematically and completed.
In particular, at present it remains an open question if among the infinitely many parameters of the 
embedding tensor --- combining higher-dimensional fluxes, torsion, etc. --- there remain
infinitely many inequivalent solutions of the quadratic constraint~(\ref{Quadratic3}).
Likewise, it will be interesting to analyze the possible higher-dimensional origin of
higher-charge components of the embedding tensor in the various gradings.

Finally, we have seen in this paper and in particular in the examples discussed,
how the algebraic structures exhibited in higher-dimensional maximal gaugings are naturally 
embedded into infinite-dimensional representations of the affine algebra $\mathfrak{e}_{9(9)}$. 
E.g.\ Figure~\ref{Figure:theta} shows how the general formulas of this paper 
can reproduce in particular all the properties and constraints 
of maximal three-dimensional gaugings.
It is moreover interesting to note that reducing in dimensions,
the two-dimensional theory is the first one in which the 
global (and subsequently gauged) symmetry  $\mathfrak{e}_{d(d)}$ combines 
--- via the central extension of $\mathfrak{e}_{9(9)}$ --- an action on the scalar matter sector with an action 
on the (non-propagating) gravitational degrees of freedom. It would be highly interesting to identify
the higher-dimensional ancestor of this mechanism.\footnote{The explicit form of~\eqref{DefABC}
suggests that in higher dimensions this corresponds to gaugings defined by an embedding tensor
of the particular form 
$\Theta_{M}{}^{A}=\eta^{AB}t_{B,M}{}^{N}\theta_{N}$, $\Theta_{M}{}^{0}=\theta_{M}$,
parametrized in terms of a $\theta_{M}$ in the conjugate vector field representation,
where the global symmetry algebra $\langle t_{A} \rangle$ has been extended by the
generator $t_{(0)}$ defining the global (on-shell) scaling symmetry of metric and $p$-forms.
These theories have not yet been considered in \cite{Nicolai:2000sc,deWit:2002vt} and belong to
the class of supergravities without actions whose nine-dimensional members have
been studied in~\cite{Bergshoeff:2002nv}.

It is also interesting to note that similar structures occur in dimensional reduction including 
the higher Kaluza-Klein modes~\cite{Hohm:2005sc}.}
From this unifying point of view, 
it would of course be of greatest interest to push the construction of gauged supergravities further down 
to even lower dimensions, embedding these structures into the
group theory of the exceptional groups ${\rm E}_{10}$~\cite{Juli85,Damour:2002cu} 
and ${\rm E}_{11}$~\cite{West:2001as,Riccioni:2007au,Bergshoeff:2007qi}.

\subsection*{Acknowledgments}
We wish to thank B.~de Wit, O.~Hohm, A.~Kleinschmidt, H.~Nicolai, M.~Ro\v cek, I.~Runkel,  S.~Sch\"afer-Nameki,
and M.~Trigiante for very helpful comments and discussions.

\begin{appendix}

\section{The algebra $\mathfrak{G}$ -- useful relations}
\label{App:algebra}

The algebra $\mathfrak{G}$ extending the affine algebra $\hat{\mathfrak{g}}$ 
by $L_{1}$ is generated by generators $T_{\alpha,m}$, $L_{1}$, $K$\,, 
with commutation relations
\bea
{}[\;T_{\alpha,m}\;,\;T_{\beta,n}\:] &=& f_{\alpha\beta}{}^{\gamma}\,T_{\gamma,m+n}+
m\,\delta_{m+n}\,\eta_{\alpha\beta}\,K
\;, \nonumber\\
{}[\,L_1,T_{\alpha,m}\,]&=& -m\,T_{\alpha,m+1}  \;,
\label{commA}
\eea
and all other commutators vanishing.
We parametrize an arbitrary algebra element as
\bea
   \Lambda &=& \Lambda^\cAa \,T_\cAa ~=~ 
   \Lambda^{\alpha,m}\, T_{\alpha,m} \, 
   +  \, \Lambda^{(1)} \, L_{1} \, + \, \Lambda^{(0)} \, K
   ~\equiv~ 
   \Lambda(w)  
   +   \Lambda^{(1)}  L_{1}  +  \Lambda^{(0)} \, K\;,
   \label{SymParamD2A}
\eea
with $\Lambda(w)\equiv \Lambda^{\alpha,m}w^{-m}\,t_{\alpha}$,
such that the commutators~(\ref{commA}) translate into
\bea
{}\leftb\Lambda,\Sigma\,\rightb &=&
[\Lambda(w),\Sigma(w)]+
\Lambda^{(1)}\partial\Sigma(w)-
\Sigma^{(1)}\partial\Lambda(w)+
K\,\Big\langle \Lambda(w)\,\partial\Sigma(w)\Big\rangle_{\!w}
\;,
\eea
and the invariant bilinear form~(\ref{invbil}) is given by
\bea
\Big(\Lambda,\Sigma\Big) &=&
{\rm tr}\,\Big\langle \Lambda(w)\,\Sigma(w)\Big\rangle_{\!w}
-\Lambda^{(1)}\Sigma^{(0)}-\Sigma^{(1)}\Lambda^{(0)}
\;.
\eea
Strictly speaking, we will consider only such elements $\Lambda\in\mathfrak{G}$
for which almost all $\{\Lambda^{\alpha,m}\,|\,m<0\}$ are equal to zero, i.e.\ for which the
power series $\Lambda(w)$ has only a finite number of positive powers.

For a general power series $f(w)=\sum_{m=-\infty}^{\infty} f_m w^{m}$
with almost all $\{f_{m}\,|\,m>0\}$ equal to zero, one proves the relation
\begin{align}
   \left \langle \frac{ f(v) } {v-w} \right \rangle_{\!v} 
          &=  \left \langle \sum_{m\ge 0}\frac{ f(v) \, w^m } {v^{m+1}} \right \rangle_{\!v}
     = \sum_{m\ge 0} f_m w^m \; .
\label{rel1}
\end{align}
Another relation that we will repeatedly make use of is
\begin{align}
   \Big\langle \Big\langle \frac{f(w,v)} {v-w} \Big\rangle_{\!v} \:\Big\rangle_{\!w} 
    - \Big\langle \Big\langle \frac{f(w,v)} {v-w} \Big\rangle_{\!w}\: \Big\rangle_{\!v}
   &= \langle f(w,w) \rangle_w \;.
   \label{rel2}   
\end{align}

\end{appendix}

{\small
%\bibliographystyle{Jopt2}
%\bibliography{refs}

\providecommand{\href}[2]{#2}\begingroup\raggedright\endgroup

}

\end{document}